\documentclass[journal]{IEEEtran}
\ifCLASSINFOpdf
\else
\fi
\usepackage{amsmath}
\usepackage{amsthm}
\usepackage{amsfonts}
\usepackage{extarrows}
\usepackage{graphicx}
\usepackage{epsfig}
\usepackage{amssymb}
\usepackage{subfigure}
\usepackage{tabularx}
\usepackage{makecell}
\usepackage{hhline}
\usepackage{booktabs}
\usepackage{harpoon}
\usepackage{balance}
\usepackage{epstopdf}
\usepackage{cases}
\usepackage{threeparttable}
\usepackage{multicol}
\usepackage{multirow}
\makeatletter
\renewcommand{\maketag@@@}[1]{\hbox{\m@th\normalsize\normalfont#1}}
\makeatother
\newcommand\relphantom[1]{\mathrel{\phantom{#1}}}
\begin{document}
\title{Symplectic Wigner Distribution in the Linear Canonical Transform Domain: Theory and Application}

\author{Yangfan~He and Zhichao~Zhang,~\IEEEmembership{Member,~IEEE}
\thanks{This work was supported in part by the Foundation of Key Laboratory of System Control and Information Processing, Ministry of Education under Grant Scip20240121; in part by the Foundation of Key Laboratory of Computational Science and Application of Hainan Province under Grant JSKX202401; in part by the Foundation of Key Laboratory of Numerical Simulation of Sichuan Provincial Universities under Grant KLNS--2024SZFZ005; and in part by the Startup Foundation for Introducing Talent of Nanjing Institute of Technology under Grant YKJ202214. \emph{(Corresponding author: Zhichao~Zhang.)}}
\thanks{Yangfan~He is with the School of Communication and Artificial Intelligence, School of Integrated Circuits, Nanjing Institute of Technology, Nanjing 211167, China (e-mail: yangf\_he@163.com).}
\thanks{Zhichao~Zhang is with the School of Mathematics and Statistics, Nanjing University of Information Science and Technology, Nanjing 210044, China, with the Key Laboratory of System Control and Information Processing, Ministry of Education, Shanghai Jiao Tong University, Shanghai 200240, China, with the Key Laboratory of Computational Science and Application of Hainan Province, Hainan Normal University, Haikou 571158, China, and also with the Key Laboratory of Numerical Simulation of Sichuan Provincial Universities, School of Mathematics and Information Sciences, Neijiang Normal University, Neijiang 641000, China (e-mail: zzc910731@163.com).}}

\markboth{IEEE TRANSACTIONS ON SIGNAL PROCESSING}
{Shell \MakeLowercase{\textit{et al.}}: Bare Demo of IEEEtran.cls for Journals}

\maketitle

\begin{abstract}
This paper devotes to combine the chirp basis function transformation and symplectic coordinates transformation to yield a novel Wigner distribution (WD) associated with the linear canonical transform (LCT), named as the symplectic WD in the LCT domain (SWDL). It incorporates the merits of the symplectic WD (SWD) and the WD in the LCT domain (WDL), achieving stronger capability in the linear frequency-modulated (LFM) signal frequency rate feature extraction while maintaining the same level of computational complexity. Some essential properties of the SWDL are derived, including marginal distributions, energy conservations, unique reconstruction, Moyal formula, complex conjugate symmetry, time reversal symmetry, scaling property, time translation property, frequency modulation property, and time translation and frequency modulation property. Heisenberg's uncertainty principles of the SWDL are formulated, giving rise to three kinds of lower bounds attainable respectively by Gaussian enveloped complex exponential signal, Gaussian signal and Gaussian enveloped chirp signal. The optimal symplectic matrices corresponding to the highest time-frequency resolution are generated by solving the lower bound optimization (minimization) problem. The time-frequency resolution of the SWDL is compared with those of the SWD and WDL to demonstrate its superiority in LFM signals time-frequency energy concentration. A synthesis example is also carried out to verify the feasibility and reliability of the theoretical analysis.
\end{abstract}


\IEEEpeerreviewmaketitle

\section{Introduction}
\IEEEPARstart{W}{igner} distribution (WD) [1], also known as Wigner-Ville distribution, is one of the most fundamental bilinear time-frequency distributions [2]. It plays an important role in time-frequency analysis of non-stationary signals whose frequencies change over time. Particularly, it is able to effectively extract the frequency rate characteristic of linear frequency-modulated (LFM) signals [3]--[5], which are frequently encountered in radar, communications, sonar, biomedicine and vibration engineering. It becomes therefore interesting and meaningful to explore how to enhance the LFM signal frequency rate feature extraction capability of the WD.\\
\indent Fractional domain time-frequency transforms break through the frequency domain analysis limitation of the traditional Fourier transform (FT), offering the fractional domain signal characterization that is between the time domain and frequency domain [6], [7]. This opens up a new way to the investigation of non-stationary signals. The linear canonical transform (LCT) [8], [9] is one of the most representative fractional domain time-frequency transforms. Its high-dimensional form is known as the metaplectic transform (MT) [10], [11], which can deal with multi-dimensional non-stationary signals especially for two-dimensional (2D) images and 3D point clouds. This paper focuses mainly on the 1D case. The LCT includes particular cases some well-known signal processing tools, including the FT, fractional Fourier transform [12], [13], Fresnel transform, Lorentz transform, and scaling and chirp multiplication operations. Thanks to its chirp basis function, the LCT is suitable to analyze LFM signals. It is essentially a 2D geometric transformation in time-frequency plane, which can be decomposed into rotation, shear and scaling. This enables the LCT to have enough flexibility and freedom in solving LFM signal processing problems, such as filtering [14], reconstruction [15], detection [16], etc.\\
\indent To improve LFM signal analysis performance, there are many attempts to integrate the WD deeply with the LCT, giving birth to a large number of WD's variants associated with the LCT. The first result in this field is the affine characteristic type of WD (ACWD) introduced by Pei \emph{et al.} [17]. It turns out that the ACWD is an effective multi-component signal separation tool. Moreover, Wei \emph{et al.} [18] explored some main properties of the ACWD and applied it to analyze first-order optical systems. In 2012, Bai \emph{et al.} [19] replaced the sinusoidal basis function with the chirp basis function, giving rise to the WD in the LCT domain (WDL). The simulated results state that the WDL achieves better detection performance on noisy LFM signals than the WD. Song \emph{et al.} [20] extended the WDL into a generalized form by using a four-term product type of instantaneous auto-correlation function, not only inheriting the capability of the WDL for LFM signal detection but also forming its specific superiority in quadratic frequency-modulated signal detection. By using one LCT type of instantaneous cross-correlation function (ICF), Zhang [21] extended the WDL into the so-called ICF type of WD (ICFWD). The ICFWD was further extended by Zhang \emph{et al.} [22] into a closed-form one (CICFWD) by applying two different LCTs type of ICF. Note that applying two coincident LCTs type of ICF generates none other than the ACWD. Compared with the WDL, both the ICFWD and CICFWD have to sacrifice the parametric complexity and computational complexity to achieve better detection performance on noisy LFM signals [23]--[25]. All of the above-mentioned variants belong to the family of chirp basis function transformation. In addition, the frequency-modulated type of WD (FMWD) proposed respectively by Zhang [26] and Xin \emph{et al.} [27], and the convolution representation type of WD (CRWD) proposed by Zhang [28], belong also to this family.\\
\indent The family of symplectic coordinates transformation is another class of variants that employ the general symplectic coordinates transformation in substituting the permanent scale (i.e., the scale $1/2$) coordinates transformation found in the WD or the single scale (i.e., the scale $\tau$) coordinates transformation found in the $\tau$-WD, explored formerly by Janssen [29], Wong [30] and Shubin [31], and more recently by Boggiatto \emph{et al.} [32], Luef \emph{et al.} [33] and Cordero \emph{et al.} [34]. The most concise and general form in this field is the symplectic WD (SWD) introduced by Zhang \emph{et al.} [35]. It is none other than a kind of specific matrix WD proposed by Bayer \emph{et al.} [36] and Cordero \emph{et al.} [37], where the matrix is a symplectic matrix. It is also a special case of the non-cross metaplectic WD (NCMWD) formulated by Zhang \emph{et al.} [38] and metaplectic WDs (MWDs) given by Cordero \emph{et al.} [39]--[42] and Gr\"{o}chenig \emph{et al.} [43]. The derived Heisenberg's uncertainty principles of the SWD and the related lower bound minimization analysis results demonstrate that the SWD exhibits higher time-frequency resolution than the WD, while maintaining the same level of computational complexity.\\
\indent To sum up, the WDL and SWD are typical representative variants in families of chirp basis function transformation and symplectic coordinates transformation, respectively. These two variants have their own advantages in LFM signal processing as compared with the WD, without sacrificing the computational complexity. The main purpose of this paper is to combine them organically, achieving stronger capability in the LFM signal frequency rate feature extraction. The main contributions of this paper are summarized as follows:
\begin{itemize}

    \item We conduct an organic integration of the SWD and WDL, giving birth to the definition of the so-called SWD in the LCT domain (SWDL).

    \item We obtain many useful and effective properties of the SWDL, which are generalizations of those of the SWD, WDL and WD.

    \item We derive the computational complexity of the SWDL, which is the same as those of the SWD, WDL and WD.

    \item We establish Heisenberg's uncertainty principles of the SWDL, based on which we solve the symplectic matrices selection issue, indicating that the SWDL outperforms the SWD, WDL and WD in LFM signals time-frequency energy concentration.

\end{itemize}

\noindent The main differences and connections between the current work and the previous ones are concluded below:
\begin{itemize}

    \item The SWDL differs essentially from the existing WD's variants associated with the LCT, including the ACWD, WDL, ICFWD, CICFWD, FMWD, CRWD and SWD.

    \item The SWDL includes particular cases the SWD, WDL and WD.

    \item The SWDL can be regarded as the 1D case of the NCMWD and a special case of MWDs, that deserves to be studied on its own.

\end{itemize}

\indent The remainder of this paper is structured as follows. In Section~\ref{sec:2}, we collect some preparatory works about the metaplectic representation, the SWD and the WDL. In Section~\ref{sec:3}, we establish fundamental theories of the SWDL, including the definition, properties and computational complexity. In Section~\ref{sec:4}, we formulate a number of Heisenberg's uncertainty inequalities of the SWDL. In Section~\ref{sec:5}, we investigate the selection of symplectic matrices with which the SWDL achieves the highest time-frequency resolution and time-frequency superresolution, especially on LFM signals. In Section~\ref{sec:6}, we provide simulations to verify the theoretical result. In Section~\ref{sec:7}, we draw a conclusion.
\section{Preliminaries}\label{sec:2}
\indent In this section, we recall some necessary background and notation on the metaplectic representation, the SWD and the WDL.
\subsection{Metaplectic representation}
\indent The family of symplectic matrices is known as the symplectic group, defined by [35]
\begin{equation*}
Sp(N,\mathbb{R})=\left\{\mathcal{A}\in GL(2N,\mathbb{R}):\mathcal{A}\mathbf{L}_N\mathcal{A}^{\mathrm{T}}=\mathbf{L}_N\right\},
\end{equation*}
(this definition is not the same as that introduced in [34], but they are equivalent by interchanging the symplectic matrix $\mathcal{A}$ and its transpose $\mathcal{A}^{\mathrm{T}}$), where $GL(2N,\mathbb{R})$ denotes the linear group of $2N\times 2N$ invertible matrices, the superscript $\mathrm{T}$ denotes the transpose operator, and $\mathbf{L}_N:=\begin{pmatrix}\mathbf{0}_{N\times N}&\mathbf{I}_{N\times N}\\-\mathbf{I}_{N\times N}&\mathbf{0}_{N\times N}\end{pmatrix}$, (here $\mathbf{I}_{N\times N}$ and $\mathbf{0}_{N\times N}$ are the $N\times N$ identity matrix and null matrix, respectively).\\
\indent The metaplectic representation, also known as the Shale-Weil representation, is a unitary representation of the (double cover of the) symplectic group $Sp(N,\mathbb{R})$ on $L^2(\mathbb{R}^N)$. It arises as intertwining operator between the standard Schr\"{o}dinger representation $\rho$ of the Heisenberg group $\mathbb{H}^N$ and the representation that is obtained from it by composing $\rho$ with the action of $Sp(N,\mathbb{R})$ by automorphisms on $\mathbb{H}^N$ [11], [38].\\
\indent Given $\mathcal{A}\in Sp(N,\mathbb{R})$ with the $N\times N$ block decomposition $\begin{pmatrix}\mathbf{A}&\mathbf{B}\\\mathbf{C}&\mathbf{D}\end{pmatrix}$, it follows from $\mathcal{A}\mathbf{L}_N\mathcal{A}^{\mathrm{T}}=\mathbf{L}_N$ that the blocks $\mathbf{A},\mathbf{B},\mathbf{C},\mathbf{D}$ must satisfy
\begin{subnumcases}
{}
\mathbf{A}\mathbf{B}^{\mathrm{T}}=\mathbf{B}\mathbf{A}^{\mathrm{T}},\\
\mathbf{C}\mathbf{D}^{\mathrm{T}}=\mathbf{D}\mathbf{C}^{\mathrm{T}},\\
\mathbf{A}\mathbf{D}^{\mathrm{T}}-\mathbf{B}\mathbf{C}^{\mathrm{T}}=\mathbf{I}_{N\times N}.
\end{subnumcases}
The MT of the function $f\in L^2(\mathbb{R}^N)$ associated with the symplectic matrix $\mathcal{A}=\begin{pmatrix}\mathbf{A}&\mathbf{B}\\\mathbf{C}&\mathbf{D}\end{pmatrix}\in Sp(N,\mathbb{R})$ with $\mathrm{det}(\mathbf{B})\neq0$ is defined as [38]
\begin{equation*}
\mathcal{M}^{\mathcal{A}}f(\mathbf{u})=\int_{\mathbb{R}^N}f(\mathbf{t})\mathcal{MK}^{\mathcal{A}}(\mathbf{u},\mathbf{t})\mathrm{d}\mathbf{t},
\end{equation*}
where
\begin{align*}
\mathcal{MK}^{\mathcal{A}}(\mathbf{u},\mathbf{t})=&\frac{1}{\sqrt{-\mathrm{det}(2\pi\mathbf{B})}}\notag\\
&\times\mathrm{e}^{\mathrm{j}\left(\frac{\mathbf{u}\mathbf{D}\mathbf{B}^{-1}\mathbf{u}^{\mathrm{T}}}{2}-\mathbf{t}\mathbf{B}^{-1}\mathbf{u}^{\mathrm{T}}+\frac{\mathbf{t}\mathbf{B}^{-1}\mathbf{A}\mathbf{t}^{\mathrm{T}}}{2}\right)}.
\end{align*}
\indent In this paper, we focus mainly on the symplectic group $Sp(N,\mathbb{R})$ in the case of $N=1$, i.e., $Sp(1,\mathbb{R})$. For a $2\times2$ symplectic matrix $\mathcal{A}=\begin{pmatrix}a&b\\c&d\end{pmatrix}\in Sp(1,\mathbb{R})$, the relations (1a) and (1b) hold naturally and the relation (1c) turns into $\mathrm{det}(\mathcal{A})=ad-bc=1$. Thus, the symplectic group $Sp(1,\mathbb{R})$ is given by [35]
\begin{equation*}
Sp(1,\mathbb{R})=\left\{\mathcal{A}=\begin{pmatrix}a&b\\c&d\end{pmatrix}\in GL(2,\mathbb{R}):ad-bc=1\right\}.
\end{equation*}
In that case, the MT is none other than the LCT with $b\neq0$, defined by [8]
\begin{equation*}
\mathcal{L}^{\mathcal{A}}f(u)=\int_{\mathbb{R}}f(t)\mathcal{LK}^{\mathcal{A}}(u,t)\mathrm{d}t,
\end{equation*}
where
\begin{equation*}
\mathcal{LK}^{\mathcal{A}}(u,t)=\frac{1}{\sqrt{\mathrm{j}2\pi b}}\mathrm{e}^{\mathrm{j}\left(\frac{d}{2b}u^2-\frac{1}{b}ut+\frac{a}{2b}t^2\right)}.
\end{equation*}
As for $b=0$, the LCT is essentially a scaling and chirp multiplication operations, given by $\sqrt{d}\mathrm{e}^{\mathrm{j}cd/2u^2}f(du)$. The Fourier operator $\mathcal{F}$ is a special case of the linear canonical operator $\mathcal{L}^{\mathcal{A}}$ corresponding to $\mathcal{A}=\mathbf{L}_1$, i.e.,
\begin{equation*}
\mathcal{F}f(\omega)=\sqrt{\mathrm{j}}\mathcal{L}^{\mathbf{L}_1}f(\omega)=\frac{1}{\sqrt{2\pi}}\int_{\mathbb{R}}f(t)\mathrm{e}^{-\mathrm{j}\omega t}\mathrm{d}t.
\end{equation*}
The LCT is invertible because of
\begin{equation*}
f=\mathcal{L}^{\mathcal{A}^{-1}}\mathcal{L}^{\mathcal{A}}f,
\end{equation*}
where $\mathcal{A}^{-1}=\begin{pmatrix}d&-b\\-c&a\end{pmatrix}\in Sp(1,\mathbb{R})$.
\subsection{Symplectic WD (SWD)}
\indent Let a signal $f\in L^2(\mathbb{R})$ and a symplectic matrix $\mathcal{A}_1=\begin{pmatrix}a_1&b_1\\c_1&d_1\end{pmatrix}\in Sp(1,\mathbb{R})$. The SWD of the signal $f$ is defined as [35]
\begin{align*}
\mathrm{W}_{\mathcal{A}_1}f(t,\omega)=&\mathcal{F}\mathfrak{T}_{\mathcal{A}_1\mathbf{J}}(f\otimes\overline{f})(t,\omega)\notag\\
=&\frac{1}{\sqrt{2\pi}}\int_{\mathbb{R}}f(b_1t+d_1\varepsilon)\overline{f(a_1t+c_1\varepsilon)}\mathrm{e}^{-\mathrm{j}\omega\varepsilon}\mathrm{d}\varepsilon,
\end{align*}
where the tensor product $\otimes$ and the coordinates operator $\mathfrak{T}_{\mathcal{A}_1\mathbf{J}}$ with the symplectic matrix $\mathcal{A}_1$ are given by $(f\otimes\overline{g})(t,\varepsilon):=f(t)\overline{g(\varepsilon)}$ and $\mathfrak{T}_{\mathcal{A}_1\mathbf{J}}h(t,\varepsilon):=\sqrt{|\mathrm{det}(\mathcal{A}_1\mathbf{J})|}h((t,\varepsilon)\mathcal{A}_1\mathbf{J})=h(b_1t+d_1\varepsilon,a_1t+c_1\varepsilon)$, respectively, and where $\mathbf{J}:=\begin{pmatrix}0&1\\1&0\end{pmatrix}$ and the superscript --- denotes the complex conjugate operator. Here, the Fourier operator $\mathcal{F}$ is essentially a partial Fourier operator.\\
\indent Obviously, the WD is a special case of the SWD corresponding to $\mathcal{A}_1=\begin{pmatrix}1&1\\-1/2&1/2\end{pmatrix}\in Sp(1,\mathbb{R})$. Thus, the SWD can be generated by extending the specific symplectic coordinates operator $\mathfrak{T}_{\begin{pmatrix}1&1\\1/2&-1/2\end{pmatrix}}$ found in the WD to a general one $\mathfrak{T}_{\mathcal{A}_1\mathbf{J}}$.
\subsection{WD in the LCT domain (WDL)}
\indent Let a signal $f\in L^2(\mathbb{R})$, and let $\mathcal{L}^{\mathcal{A}_2}$ be a linear canonical operator associated with the symplectic matrix $\mathcal{A}_2=\begin{pmatrix}a_2&b_2\\c_2&d_2\end{pmatrix}\in Sp(1,\mathbb{R})$ with $b_2\neq0$. The WDL of the signal $f$ is defined as [19]
\begin{align*}
\mathrm{W}^{\mathcal{A}_2}f(t,u)=&\mathcal{L}^{\mathcal{A}_2}\mathfrak{T}_{\begin{pmatrix}1&1\\1/2&-1/2\end{pmatrix}}(f\otimes\overline{f})(t,u)\notag\\
=&\int_{\mathbb{R}}f\left(t+\frac{\varepsilon}{2}\right)\overline{f\left(t-\frac{\varepsilon}{2}\right)}\mathcal{LK}^{\mathcal{A}_2}(u,\varepsilon)\mathrm{d}\varepsilon.
\end{align*}
Here the linear canonical operator $\mathcal{L}^{\mathcal{A}_2}$ is essentially a partial linear canonical operator.\\
\indent The WD is also a special case of the WDL corresponding to $\mathcal{A}_2=\mathbf{L}_1\in Sp(1,\mathbb{R})$. Different essentially from the SWD, thus, the WDL can be generated by extending the Fourier operator $\mathcal{F}$ found in the WD to a linear canonical one $\mathcal{L}^{\mathcal{A}_2}$.
\section{SWD in the LCT domain (SWDL)}\label{sec:3}
\indent In this section, we formulate the definition of the SWDL and revisit it by using the LCT. Then, we derive some important and useful properties of the SWDL. We also analyze the computational complexity of the SWDL.
\subsection{Definition}
\indent Combining the SWD and WDL yields the definition of the SWDL, which is given below.\\
\indent\emph{Definition~1:} Let a signal $f\in L^2(\mathbb{R})$ and a symplectic matrix $\mathcal{A}_1=\begin{pmatrix}a_1&b_1\\c_1&d_1\end{pmatrix}\in Sp(1,\mathbb{R})$, and let $\mathcal{L}^{\mathcal{A}_2}$ be a linear canonical operator associated with the symplectic matrix $\mathcal{A}_2=\begin{pmatrix}a_2&b_2\\c_2&d_2\end{pmatrix}\in Sp(1,\mathbb{R})$ with $b_2\neq0$. The SWDL of the signal $f$ is defined as
\begin{align}\label{eq2}
\mathrm{W}_{\mathcal{A}_1}^{\mathcal{A}_2}f(t,u)=&\mathcal{L}^{\mathcal{A}_2}\mathfrak{T}_{\mathcal{A}_1\mathbf{J}}(f\otimes\overline{f})(t,u)\notag\\
=&\int_{\mathbb{R}}f(b_1t+d_1\varepsilon)\overline{f(a_1t+c_1\varepsilon)}\mathcal{LK}^{\mathcal{A}_2}(u,\varepsilon)\mathrm{d}\varepsilon.
\end{align}
\indent The SWDL is an organic integration of the SWD and WDL, and includes them as particular cases, as shown in Table~\ref{tab:1}.
\begin{table}[htbp]
\centering
\caption{\label{tab:1}Some special cases of the SWDL}
\footnotesize
\begin{tabular}{ccc}
\specialrule{0.1em}{4pt}{4pt}
$\mathcal{A}_1=\begin{pmatrix}a_1&b_1\\c_1&d_1\end{pmatrix}$ & $\mathcal{A}_2=\begin{pmatrix}a_2&b_2\\c_2&d_2\end{pmatrix}$ & SWDL\\
\specialrule{0.1em}{4pt}{4pt}
$\mathcal{A}_1$ & $\mathbf{L}_1$ & SWD\\
\specialrule{0em}{4pt}{4pt}
$\begin{pmatrix}1&1\\-1/2&1/2\end{pmatrix}$ & $\mathcal{A}_2$ & WDL\\
\specialrule{0em}{4pt}{4pt}
$\begin{pmatrix}1&1\\-1/2&1/2\end{pmatrix}$ & $\mathbf{L}_1$ & WD\\
\specialrule{0.1em}{4pt}{4pt}
\end{tabular}
\end{table}
\subsection{Definition revisited in terms of the LCT}
\indent In addition to the definition of the SWDL in the form of the tensor product of the original signal, there exists an equivalent form of the tensor product of the LCTed signals.\\
\indent\emph{Lemma~1:} Let $\mathrm{W}_{\mathcal{A}_1}^{\mathcal{A}_2}f$ be the SWDL of $f\in L^2(\mathbb{R})$ associated with the symplectic matrices $\mathcal{A}_1=\begin{pmatrix}a_1&b_1\\c_1&d_1\end{pmatrix}\in Sp(1,\mathbb{R})$ with $a_1,b_1,c_1,d_1\neq0$ and $\mathcal{A}_2=\begin{pmatrix}a_2&b_2\\c_2&d_2\end{pmatrix}\in Sp(1,\mathbb{R})$ with $b_2\neq0$, and let $\mathcal{L}^{\mathcal{A}_3},\mathcal{L}^{\mathcal{A}_4}$ be two linear canonical operators associated with the symplectic matrices $\mathcal{A}_3=\begin{pmatrix}a_2a_1&b_2d_1\\c_2/d_1&d_2/a_1\end{pmatrix},\mathcal{A}_4=\begin{pmatrix}a_2b_1&b_2c_1\\c_2/c_1&d_2/b_1\end{pmatrix}\in Sp(1,\mathbb{R})$, respectively. The SWDL can be rewritten in terms of the LCT as
\begin{align}\label{eq3}
\mathrm{W}_{\mathcal{A}_1}^{\mathcal{A}_2}f(t,u)=\chi(u,t)\mathcal{F}\mathfrak{T}_{\mathbf{M}}\left(\widehat{\mathcal{L}^{\mathcal{A}_3}f}\otimes\overline{\widetilde{\mathcal{L}^{\mathcal{A}_4}f}}\right)\left(u,\frac{t}{b_2c_1d_1}\right),
\end{align}
where $\chi(u,t):=\frac{\mathrm{e}^{\mathrm{j}\left(\frac{d_2}{2b_2}u^2+\frac{a_1}{b_2c_1}ut+\frac{a_2a_1b_1}{2b_2c_1d_1}t^2\right)}}{\sqrt{\mathrm{j}b_2c_1d_1}}$, $\mathbf{M}:=\begin{pmatrix}0&-1\\1&1\end{pmatrix}$, $\widehat{\mathcal{L}^{\mathcal{A}_3}f}(u):=\mathrm{e}^{-\mathrm{j}\frac{d_2}{2b_2a_1d_1}u^2}\mathcal{L}^{\mathcal{A}_3}f(u)$ and $\widetilde{\mathcal{L}^{\mathcal{A}_4}f}(v):=\mathrm{e}^{-\mathrm{j}\frac{d_2}{2b_2b_1c_1}v^2}\mathcal{L}^{\mathcal{A}_4}f(v)$.\\
\indent\emph{Proof:} By using the invertibility of the LCT, we have
\begin{align*}
f(b_1t+d_1\varepsilon)=&\mathcal{L}^{\mathcal{A}_3^{-1}}\mathcal{L}^{\mathcal{A}_3}f(b_1t+d_1\varepsilon)\notag\\
=&\int_{\mathbb{R}}\mathcal{L}^{\mathcal{A}_3}f(v)\mathcal{LK}^{\mathcal{A}_3^{-1}}(b_1t+d_1\varepsilon,v)\mathrm{d}v,
\end{align*}
\begin{align*}
f(a_1t+c_1\varepsilon)=&\mathcal{L}^{\mathcal{A}_4^{-1}}\mathcal{L}^{\mathcal{A}_4}f(a_1t+c_1\varepsilon)\notag\\
=&\int_{\mathbb{R}}\mathcal{L}^{\mathcal{A}_4}f(\mu)\mathcal{LK}^{\mathcal{A}_4^{-1}}(a_1t+c_1\varepsilon,\mu)\mathrm{d}\mu,
\end{align*}
based on which the SWDL becomes
\begin{align}\label{eq4}
\mathrm{W}_{\mathcal{A}_1}^{\mathcal{A}_2}f(t,u)=&\int_{\mathbb{R}^3}\mathcal{L}^{\mathcal{A}_3}f(v)\overline{\mathcal{L}^{\mathcal{A}_4}f(\mu)}\mathcal{LK}^{\mathcal{A}_2}(u,\varepsilon)\notag\\
&\times\mathcal{LK}^{\mathcal{A}_3^{-1}}(b_1t+d_1\varepsilon,v)\overline{\mathcal{LK}^{\mathcal{A}_4^{-1}}(a_1t+c_1\varepsilon,\mu)}\notag\\
&\times\mathrm{d}\varepsilon\mathrm{d}\mu\mathrm{d}v.
\end{align}
Thanks to
\begin{align*}
&\mathcal{LK}^{\mathcal{A}_2}(u,\varepsilon)\mathcal{LK}^{\mathcal{A}_3^{-1}}(b_1t+d_1\varepsilon,v)\overline{\mathcal{LK}^{\mathcal{A}_4^{-1}}(a_1t+c_1\varepsilon,\mu)}\notag\\
&=\kappa(t,u;v,\mu)\frac{\mathrm{e}^{-\mathrm{j}\frac{1}{b_2}(u-v+\mu)\varepsilon}}{2\pi|b_2|},
\end{align*}
where
\begin{align*}
\kappa(t,u;v,\mu):=&\frac{\mathrm{e}^{\mathrm{j}\left(\frac{d_2}{2b_2}u^2+\frac{a_2a_1b_1}{2b_2c_1d_1}t^2\right)}}{\sqrt{\mathrm{j}2\pi b_2c_1d_1}}\mathrm{e}^{-\mathrm{j}\left(\frac{d_2}{2b_2a_1d_1}v^2-\frac{d_2}{2b_2b_1c_1}\mu^2\right)}\notag\\
&\times\mathrm{e}^{\mathrm{j}\left(\frac{b_1}{b_2d_1}tv-\frac{a_1}{b_2c_1}t\mu\right)},
\end{align*}
it follows that
\begin{align*}
&\int_{\mathbb{R}}\mathcal{LK}^{\mathcal{A}_2}(u,\varepsilon)\mathcal{LK}^{\mathcal{A}_3^{-1}}(b_1t+d_1\varepsilon,v)\overline{\mathcal{LK}^{\mathcal{A}_4^{-1}}(a_1t+c_1\varepsilon,\mu)}\mathrm{d}\varepsilon\notag\\
&=\kappa(t,u;v,\mu)\left(\frac{1}{2\pi|b_2|}\int_{\mathbb{R}}\mathrm{e}^{-\mathrm{j}\frac{1}{b_2}(u-v+\mu)\varepsilon}\mathrm{d}\varepsilon\right)\notag\\
&=\kappa(t,u;v,\mu)\delta(u-v+\mu),
\end{align*}
where $\delta$ denotes the Dirac delta operator. According to the sifting property of Dirac delta functions, Eq.~\eqref{eq4} simplifies to
\begin{align*}
\mathrm{W}_{\mathcal{A}_1}^{\mathcal{A}_2}f(t,u)=&\int_{\mathbb{R}^2}\mathcal{L}^{\mathcal{A}_3}f(v)\overline{\mathcal{L}^{\mathcal{A}_4}f(\mu)}\kappa(t,u;v,\mu)\notag\\
&\times\delta(u-v+\mu)\mathrm{d}\mu\mathrm{d}v\notag\\
=&\int_{\mathbb{R}}\mathcal{L}^{\mathcal{A}_3}f(v)\overline{\mathcal{L}^{\mathcal{A}_4}f(v-u)}\kappa(t,u;v,v-u)\mathrm{d}v\notag\\
=&\frac{\mathrm{e}^{\mathrm{j}\left(\frac{d_2}{2b_2}u^2+\frac{a_1}{b_2c_1}ut+\frac{a_2a_1b_1}{2b_2c_1d_1}t^2\right)}}{\sqrt{\mathrm{j}2\pi b_2c_1d_1}}\notag\\
&\times\int_{\mathbb{R}}\mathrm{e}^{-\mathrm{j}\frac{d_2}{2b_2a_1d_1}v^2}\mathcal{L}^{\mathcal{A}_3}f(v)\notag\\
&\times\overline{\mathrm{e}^{-\mathrm{j}\frac{d_2}{2b_2b_1c_1}(v-u)^2}\mathcal{L}^{\mathcal{A}_4}f(v-u)}\mathrm{e}^{-\mathrm{j}\frac{1}{b_2c_1d_1}tv}\mathrm{d}v.
\end{align*}
Let us denote $\frac{\mathrm{e}^{\mathrm{j}\left(\frac{d_2}{2b_2}u^2+\frac{a_1}{b_2c_1}ut+\frac{a_2a_1b_1}{2b_2c_1d_1}t^2\right)}}{\sqrt{\mathrm{j}b_2c_1d_1}}\triangleq\chi(u,t)$, $\mathrm{e}^{-\mathrm{j}\frac{d_2}{2b_2a_1d_1}u^2}\mathcal{L}^{\mathcal{A}_3}f(u)\triangleq\widehat{\mathcal{L}^{\mathcal{A}_3}f}(u)$ and $\mathrm{e}^{-\mathrm{j}\frac{d_2}{2b_2b_1c_1}v^2}\mathcal{L}^{\mathcal{A}_4}f(v)\triangleq\widetilde{\mathcal{L}^{\mathcal{A}_4}f}(v)$, we arrive the conclusion.\qed\\
\indent The above Lemma indicates that the SWDL has an equivalent definition in the form of the tensor product of LCTed signals $ \widehat{\mathcal{L}^{\mathcal{A}_3}f}$ and $\widetilde{\mathcal{L}^{\mathcal{A}_4}f}$.
\subsection{Main properties}
\indent The essential properties of the SWDL are generalizations of those of the WD [2], including marginal distributions, energy conservations, unique reconstruction, Moyal formula, complex conjugate symmetry, time reversal symmetry, scaling property, time translation property, frequency modulation property, and time translation and frequency modulation property.\\
\indent 1. Marginal distributions\\
\indent There are four types of marginal distributions.\\
\indent (i) Time marginal distribution: The integral of the product of the SWDL and $\mathcal{LK}^{\mathcal{A}_2^{-1}}(0,u)$ with respect to the linear canonical frequency variable $u$ reads
\begin{equation*}
\int_{\mathbb{R}}\mathrm{W}_{\mathcal{A}_1}^{\mathcal{A}_2}f(t,u)\mathcal{LK}^{\mathcal{A}_2^{-1}}(0,u)\mathrm{d}u=f(b_1t)\overline{f(a_1t)}.
\end{equation*}
\indent (ii) Linear canonical frequency marginal distribution: The integral of the product of the SWDL and $\mathcal{LK}^{\mathcal{A}_5^{-1}}(t,0)$ with respect to the time variable $t$ reads
\begin{equation*}
\int_{\mathbb{R}}\mathrm{W}_{\mathcal{A}_1}^{\mathcal{A}_2}f(t,u)\mathcal{LK}^{\mathcal{A}_5^{-1}}(t,0)\mathrm{d}t=\mathcal{L}^{\mathcal{A}_3}f(a_1d_1u)\overline{\mathcal{L}^{\mathcal{A}_4}f(b_1c_1u)},
\end{equation*}
where the symplectic matrix $\mathcal{A}_5=\begin{pmatrix}a_2a_1b_1&b_2c_1d_1\\c_2/(c_1d_1)&d_2/(a_1b_1)\end{pmatrix}\in Sp(1,\mathbb{R})$.\\
\indent (iii) Time delay marginal distribution: The SWDL with $t=0$ reads
\begin{equation*}
\mathrm{W}_{\mathcal{A}_1}^{\mathcal{A}_2}f(0,u)=\int_{\mathbb{R}}f(d_1\varepsilon)\overline{f(c_1\varepsilon)}\mathcal{LK}^{\mathcal{A}_2}(u,\varepsilon)\mathrm{d}\varepsilon.
\end{equation*}
\indent (iv) Linear canonical frequency shift marginal distribution: The SWDL with $u=0$ reads
\begin{equation*}
\mathrm{W}_{\mathcal{A}_1}^{\mathcal{A}_2}f(t,0)=\int_{\mathbb{R}}f(b_1t+d_1\varepsilon)\overline{f(a_1t+c_1\varepsilon)}\mathcal{LK}^{\mathcal{A}_2}(0,\varepsilon)\mathrm{d}\varepsilon.
\end{equation*}
\indent 2. Energy conservations\\
\indent Let $\lVert\cdot\rVert_2$ denote the $L^2$-norm. There are three types of energy conservations.\\
\indent (i) Time marginal distribution based energy conservation: Integrating on both sides of the time marginal distribution formula with respect to the time variable $t$ gives
\begin{equation*}
|b_1|\int_{\mathbb{R}^2}\mathrm{W}_{\mathcal{A}_1}^{\mathcal{A}_2}f(t,u)\mathcal{LK}^{\mathcal{A}_2^{-1}}(0,u)\mathrm{d}t\mathrm{d}u=\lVert f\rVert_2^2,
\end{equation*}
where $a_1=b_1$.\\
\indent (ii) Linear canonical frequency marginal distribution based energy conservation: Integrating on both sides of the linear canonical frequency marginal distribution formula with respect to the linear canonical frequency variable $u$ gives
\begin{equation*}
|b_1|\overline{\sqrt{c_1}}\sqrt{d_1}\int_{\mathbb{R}^2}\mathrm{W}_{\mathcal{A}_1}^{\mathcal{A}_2}f(t,u)\mathcal{LK}^{\mathcal{A}_5^{-1}}(t,0)\mathrm{d}t\mathrm{d}u=\lVert f\rVert_2^2,
\end{equation*}
where $a_1=b_1$ and $a_2=d_2=0$.\\
\indent (iii) Time delay or linear canonical frequency shift marginal distribution based energy conservation: The SWDL with $t=u=0$ reads
\begin{equation*}
|d_1|\sqrt{\mathrm{j}2\pi b_2}\mathrm{W}_{\mathcal{A}_1}^{\mathcal{A}_2}f(0,0)=\lVert f\rVert_2^2,
\end{equation*}
where $c_1=d_1$ and $a_2=0$.\\
\indent 3. Unique reconstruction\\
\indent The original signal $f$ can be recovered by its SWDL according to
\begin{equation*}
f(t)=\frac{1}{\overline{f(0)}}\int_{\mathbb{R}}\mathrm{W}_{\mathcal{A}_1}^{\mathcal{A}_2}f(-c_1t,u)\mathcal{LK}^{\mathcal{A}_2^{-1}}(a_1t,u)\mathrm{d}u
\end{equation*}
or
\begin{equation*}
f(t)=\frac{1}{\overline{f(0)}}\int_{\mathbb{R}}\overline{\mathrm{W}_{\mathcal{A}_1}^{\mathcal{A}_2}f(d_1t,u)}\overline{\mathcal{LK}^{\mathcal{A}_2^{-1}}(-b_1t,u)}\mathrm{d}u.
\end{equation*}
\indent 4. Moyal formula\\
\indent Let $\langle,\rangle$ denote the inner product operator, given by $\langle\circ,\diamond\rangle:=\int_{\mathbb{R}^2}\circ(t,u)\overline{\diamond(t,u)}\mathrm{d}t\mathrm{d}u$ for two functions $\circ,\diamond$ defined on $\mathbb{R}^2$ or $\langle\triangleleft,\triangleright\rangle:=\int_{\mathbb{R}}\triangleleft(t)\overline{\triangleright(t)}\mathrm{d}t$ for two functions $\triangleleft,\triangleright$ defined on $\mathbb{R}$. The inner product of the SWDLs of $f,g$ equals to the square modulus of the inner product of $f,g$. That is
\begin{equation*}
\left\langle\mathrm{W}_{\mathcal{A}_1}^{\mathcal{A}_2}f,\mathrm{W}_{\mathcal{A}_1}^{\mathcal{A}_2}g\right\rangle_{}=|\langle f,g\rangle|^2.
\end{equation*}
In particular for $f=g$, there is Parseval's relation of the SWDL
\begin{equation}\label{eq5}
\left\lVert\mathrm{W}_{\mathcal{A}_1}^{\mathcal{A}_2}f\right\rVert_2=\lVert f\rVert_2^2.
\end{equation}
\indent 5. Complex conjugate symmetry\\
\indent Let $\mathbf{N}:=\begin{pmatrix}-1&0\\0&1\end{pmatrix}$. The SWDL of the complex conjugated signal $\overline{f}$ associated with $\mathcal{A}_1,\mathcal{A}_2$ equals to the SWDL of the original signal $f$ associated with $\mathcal{A}_1,\mathbf{N}\mathcal{A}_2\mathbf{N}$. That is
\begin{equation*}
\mathrm{W}_{\mathcal{A}_1}^{\mathcal{A}_2}\overline{f}=\overline{\mathrm{W}_{\mathcal{A}_1}^{\mathbf{N}\mathcal{A}_2\mathbf{N}}f}.
\end{equation*}
\indent 6. Time reversal symmetry\\
\indent Let $T_R$ denote the time reversal operator, given by $T_Rf(t):=f(-t)$. The SWDL of the time reversed signal $T_Rf$ associated with $\mathcal{A}_1,\mathcal{A}_2$ equals to the SWDL of the original signal $f$ associated with $-\mathcal{A}_1,\mathcal{A}_2$. That is
\begin{equation*}
\mathrm{W}_{\mathcal{A}_1}^{\mathcal{A}_2}T_Rf=\mathrm{W}_{-\mathcal{A}_1}^{\mathcal{A}_2}f.
\end{equation*}
\indent 7. Scaling property\\
\indent Let $\mathbf{P}_{\sigma}:=\begin{pmatrix}1/\sigma&0\\0&\sigma\end{pmatrix}$, and let $S_{\sigma}$ denote the scaling operator, given by $S_{\sigma}f(t):=\sqrt{\sigma}f(\sigma t)$. The SWDL of the scaled signal $S_{\sigma}f$ associated with $\mathcal{A}_1,\mathcal{A}_2$ equals to the SWDL of the original signal $f$ associated with $\mathcal{A}_1,\mathbf{P}_{\sigma}\mathcal{A}_2\mathbf{P}_{\sigma}$, accompanied by the scalings $S_{\sigma}$ and $S_{1/\sigma}$ on its time variable $t$ and linear canonical frequency variable $u$, respectively. That is
\begin{equation*}
\mathrm{W}_{\mathcal{A}_1}^{\mathcal{A}_2}S_{\sigma}f(t,u)=\mathrm{W}_{\mathcal{A}_1}^{\mathbf{P}_{\sigma}\mathcal{A}_2\mathbf{P}_{\sigma}}f\left(\sigma t,\frac{u}{\sigma}\right).
\end{equation*}
\indent 8. Time translation property\\
\indent Let $T_{\vartheta}$ denote the translation operator, given by $T_{\vartheta}f(t):=f(t-\vartheta)$. The SWDL of the time translational signal $T_{\vartheta}f$ associated with $\mathcal{A}_1,\mathcal{A}_2$ equals to the product of $\mathrm{e}^{-\mathrm{j}\frac{a_2c_2(a_1-b_1)^2}{2}\vartheta^2}\mathrm{e}^{\mathrm{j}c_2(a_1-b_1)u\vartheta}$ and the SWDL of the original signal $f$ associated with $\mathcal{A}_1,\mathcal{A}_2$, accompanied by the translations $T_{(d_1-c_1)\vartheta}$ and $T_{a_2(a_1-b_1)\vartheta}$ on its time variable $t$ and linear canonical frequency variable $u$, respectively. That is
\begin{align*}
\mathrm{W}_{\mathcal{A}_1}^{\mathcal{A}_2}T_{\vartheta}f(t,u)&=\mathrm{e}^{-\mathrm{j}\frac{a_2c_2(a_1-b_1)^2}{2}\vartheta^2}\mathrm{e}^{\mathrm{j}c_2(a_1-b_1)u\vartheta}\notag\\
&\times\mathrm{W}_{\mathcal{A}_1}^{\mathcal{A}_2}f(t-(d_1-c_1)\vartheta,u-a_2(a_1-b_1)\vartheta).
\end{align*}
\indent 9. Frequency modulation property\\
\indent Let $M_{\xi}$ denote the modulation operator, given by $M_{\xi}f(t):=\mathrm{e}^{\mathrm{j}\xi t}f(t)$. The SWDL of the frequency modulated signal $M_{\xi}f$ associated with $\mathcal{A}_1,\mathcal{A}_2$ equals to the product of $\mathrm{e}^{-\mathrm{j}\frac{b_2d_2(d_1-c_1)^2}{2}\xi^2}\mathrm{e}^{\mathrm{j}d_2(d_1-c_1)u\xi}\mathrm{e}^{\mathrm{j}(b_1-a_1)t\xi}$ and the SWDL of the original signal $f$ associated with $\mathcal{A}_1,\mathcal{A}_2$, accompanied by the translation $T_{b_2(d_1-c_1)\xi}$ on its linear canonical frequency variable $u$. That is
\begin{align*}
\mathrm{W}_{\mathcal{A}_1}^{\mathcal{A}_2}M_{\xi}f(t,u)=&\mathrm{e}^{-\mathrm{j}\frac{b_2d_2(d_1-c_1)^2}{2}\xi^2}\mathrm{e}^{\mathrm{j}d_2(d_1-c_1)u\xi}\mathrm{e}^{-\mathrm{j}(a_1-b_1)t\xi}\notag\\
&\times\mathrm{W}_{\mathcal{A}_1}^{\mathcal{A}_2}f(t,u-b_2(d_1-c_1)\xi).
\end{align*}
\indent 10. Time translation and frequency modulation property\\
\indent The SWDL of the time translational and frequency modulated signal $M_{\xi}T_{\vartheta}f$ or $T_{\vartheta}M_{\xi}f$, given by $M_{\xi}T_{\vartheta}f(t):=T_{\vartheta}M_{\xi}f(t):=\mathrm{e}^{\mathrm{j}\xi(t-\vartheta)}f(t-\vartheta)$, equals to the product of $\mathrm{e}^{-\mathrm{j}\frac{b_2d_2(d_1-c_1)^2}{2}\xi^2}\mathrm{e}^{\mathrm{j}d_2(d_1-c_1)u\xi}\mathrm{e}^{-\mathrm{j}(a_1-b_1)t\xi}\mathrm{e}^{-\mathrm{j}\frac{a_2c_2(a_1-b_1)^2}{2}\vartheta^2}$ $\mathrm{e}^{\mathrm{j}c_2(a_1-b_1)(u-b_2(d_1-c_1)\xi)\vartheta}$ and the SWDL of the original signal $f$ associated with $\mathcal{A}_1,\mathcal{A}_2$, accompanied by the translations $T_{(d_1-c_1)\vartheta}$ and $T_{a_2(a_1-b_1)\vartheta+b_2(d_1-c_1)\xi}$ on its time variable $t$ and linear canonical frequency variable $u$, respectively. That is
\begin{align*}
&\mathrm{W}_{\mathcal{A}_1}^{\mathcal{A}_2}M_{\xi}T_{\vartheta}f(t,u)\notag\\
&=\mathrm{W}_{\mathcal{A}_1}^{\mathcal{A}_2}T_{\vartheta}M_{\xi}f(t,u)\notag\\
&=\mathrm{e}^{-\mathrm{j}\frac{b_2d_2(d_1-c_1)^2}{2}\xi^2}\mathrm{e}^{\mathrm{j}d_2(d_1-c_1)u\xi}\mathrm{e}^{-\mathrm{j}(a_1-b_1)t\xi}\notag\\
&\times\mathrm{e}^{-\mathrm{j}\frac{a_2c_2(a_1-b_1)^2}{2}\vartheta^2}\mathrm{e}^{\mathrm{j}c_2(a_1-b_1)(u-b_2(d_1-c_1)\xi)\vartheta}\notag\\
&\times\mathrm{W}_{\mathcal{A}_1}^{\mathcal{A}_2}f(t-(d_1-c_1)\vartheta,u-a_2(a_1-b_1)\vartheta-b_2(d_1-c_1)\xi).
\end{align*}
\indent The proofs of the above properties are similar to those of properties of the WD, and then, they are omitted. See Table~\ref{tab:2} for a summary of these properties. Note that they reduce to properties of the WD for $\mathcal{A}_1=\begin{pmatrix}1&1\\-1/2&1/2\end{pmatrix}$ and $\mathcal{A}_2=\mathbf{L}_1$, properties of the SWD for $\mathcal{A}_1$ and $\mathcal{A}_2=\mathbf{L}_1$, and properties of the WDL for $\mathcal{A}_1=\begin{pmatrix}1&1\\-1/2&1/2\end{pmatrix}$ and $\mathcal{A}_2$.
\begin{table*}[htbp]
\centering
\begin{threeparttable}
\caption{\label{tab:2}Main properties of the SWDL}
\footnotesize
\begin{tabular}{cc}
\specialrule{0.1em}{4pt}{4pt}
\multicolumn{1}{c}{Description}
&\multicolumn{1}{c}{Mathematical expression}\\\specialrule{0.1em}{4pt}{4pt}
Time marginal distribution & $\int_{\mathbb{R}}\mathrm{W}_{\mathcal{A}_1}^{\mathcal{A}_2}f(t,u)\mathcal{LK}^{\mathcal{A}_2^{-1}}(0,u)\mathrm{d}u=f(b_1t)\overline{f(a_1t)}$\\
\specialrule{0em}{4pt}{4pt}
Linear canonical frequency marginal distribution & $\int_{\mathbb{R}}\mathrm{W}_{\mathcal{A}_1}^{\mathcal{A}_2}f(t,u)\mathcal{LK}^{\mathcal{A}_5^{-1}}(t,0)\mathrm{d}t=\mathcal{L}^{\mathcal{A}_3}f(a_1d_1u)\overline{\mathcal{L}^{\mathcal{A}_4}f(b_1c_1u)}$\\
\specialrule{0em}{4pt}{4pt}
Time delay marginal distribution & $\mathrm{W}_{\mathcal{A}_1}^{\mathcal{A}_2}f(0,u)=\int_{\mathbb{R}}f(d_1\varepsilon)\overline{f(c_1\varepsilon)}\mathcal{LK}^{\mathcal{A}_2}(u,\varepsilon)\mathrm{d}\varepsilon$\\
\specialrule{0em}{4pt}{4pt}
Linear canonical frequency shift marginal distribution & $\mathrm{W}_{\mathcal{A}_1}^{\mathcal{A}_2}f(t,0)=\int_{\mathbb{R}}f(b_1t+d_1\varepsilon)\overline{f(a_1t+c_1\varepsilon)}\mathcal{LK}^{\mathcal{A}_2}(0,\varepsilon)\mathrm{d}\varepsilon$\\
\specialrule{0em}{4pt}{4pt}
Time marginal distribution & \multirow{2}{*}{$|b_1|\int_{\mathbb{R}^2}\mathrm{W}_{\mathcal{A}_1}^{\mathcal{A}_2}f(t,u)\mathcal{LK}^{\mathcal{A}_2^{-1}}(0,u)\mathrm{d}t\mathrm{d}u=\lVert f\rVert_2^2$}\\
based energy conservation\tnote{1} & \\
\specialrule{0em}{4pt}{4pt}
Linear canonical frequency marginal distribution & \multirow{2}{*}{$|b_1|\overline{\sqrt{c_1}}\sqrt{d_1}\int_{\mathbb{R}^2}\mathrm{W}_{\mathcal{A}_1}^{\mathcal{A}_2}f(t,u)\mathcal{LK}^{\mathcal{A}_5^{-1}}(t,0)\mathrm{d}t\mathrm{d}u=\lVert f\rVert_2^2$}\\
based energy conservation\tnote{2} & \\
\specialrule{0em}{4pt}{4pt}
Time delay or linear canonical frequency shift marginal distribution & \multirow{2}{*}{$|d_1|\sqrt{\mathrm{j}2\pi b_2}\mathrm{W}_{\mathcal{A}_1}^{\mathcal{A}_2}f(0,0)=\lVert f\rVert_2^2$}\\
based energy conservation\tnote{3} & \\
\specialrule{0em}{4pt}{4pt}
\multirow{3}{*}{Unique reconstruction} & $f(t)=\frac{1}{\overline{f(0)}}\int_{\mathbb{R}}\mathrm{W}_{\mathcal{A}_1}^{\mathcal{A}_2}f(-c_1t,u)\mathcal{LK}^{\mathcal{A}_2^{-1}}(a_1t,u)\mathrm{d}u,$\\
& $f(t)=\frac{1}{\overline{f(0)}}\int_{\mathbb{R}}\overline{\mathrm{W}_{\mathcal{A}_1}^{\mathcal{A}_2}f(d_1t,u)}\overline{\mathcal{LK}^{\mathcal{A}_2^{-1}}(-b_1t,u)}\mathrm{d}u$\\
\specialrule{0em}{4pt}{4pt}
Moyal formula & $\left\langle\mathrm{W}_{\mathcal{A}_1}^{\mathcal{A}_2}f,\mathrm{W}_{\mathcal{A}_1}^{\mathcal{A}_2}g\right\rangle_{}=|\langle f,g\rangle|^2$\\
\specialrule{0em}{4pt}{4pt}
Complex conjugate symmetry & \multirow{2}{*}{$\overline{\mathrm{W}_{\mathcal{A}_1}^{\mathbf{N}\mathcal{A}_2\mathbf{N}}f}$}\\
(The SWDL of $\overline{f}$: $\mathrm{W}_{\mathcal{A}_1}^{\mathcal{A}_2}\overline{f}$) & \\
\specialrule{0em}{4pt}{4pt}
Time reversal symmetry & \multirow{2}{*}{$\mathrm{W}_{-\mathcal{A}_1}^{\mathcal{A}_2}f$}\\
(The SWDL of $T_Rf$: $\mathrm{W}_{\mathcal{A}_1}^{\mathcal{A}_2}T_Rf$) & \\
\specialrule{0em}{4pt}{4pt}
Scaling property & \multirow{2}{*}{$\mathrm{W}_{\mathcal{A}_1}^{\mathbf{P}_{\sigma}\mathcal{A}_2\mathbf{P}_{\sigma}}f\left(\sigma t,\frac{u}{\sigma}\right)$}\\
(The SWDL of $S_{\sigma}f$: $\mathrm{W}_{\mathcal{A}_1}^{\mathcal{A}_2}S_{\sigma}f(t,u)$) &\\
\specialrule{0em}{4pt}{4pt}
Time translation property & $\mathrm{e}^{-\mathrm{j}\frac{a_2c_2(a_1-b_1)^2}{2}\vartheta^2}\mathrm{e}^{\mathrm{j}c_2(a_1-b_1)u\vartheta}$\\
(The SWDL of $T_{\vartheta}f$: $\mathrm{W}_{\mathcal{A}_1}^{\mathcal{A}_2}T_{\vartheta}f(t,u)$) & $\times\mathrm{W}_{\mathcal{A}_1}^{\mathcal{A}_2}f(t-(d_1-c_1)\vartheta,u-a_2(a_1-b_1)\vartheta)$\\
\specialrule{0em}{4pt}{4pt}
Frequency modulation property & $\mathrm{e}^{-\mathrm{j}\frac{b_2d_2(d_1-c_1)^2}{2}\xi^2}\mathrm{e}^{\mathrm{j}d_2(d_1-c_1)u\xi}\mathrm{e}^{-\mathrm{j}(a_1-b_1)t\xi}$\\
(The SWDL of $M_{\xi}f$: $\mathrm{W}_{\mathcal{A}_1}^{\mathcal{A}_2}M_{\xi}f(t,u)$) & $\times\mathrm{W}_{\mathcal{A}_1}^{\mathcal{A}_2}f(t,u-b_2(d_1-c_1)\xi)$\\
\specialrule{0em}{4pt}{4pt}
\multirow{2}{*}{Time translation and frequency modulation property} & $\mathrm{e}^{-\mathrm{j}\frac{b_2d_2(d_1-c_1)^2}{2}\xi^2}\mathrm{e}^{\mathrm{j}d_2(d_1-c_1)u\xi}\mathrm{e}^{-\mathrm{j}(a_1-b_1)t\xi}$\\
\multirow{2}{*}{(The SWDL of $M_{\xi}T_{\vartheta}f$ or $T_{\vartheta}M_{\xi}f$: $\mathrm{W}_{\mathcal{A}_1}^{\mathcal{A}_2}M_{\xi}T_{\vartheta}f(t,u)$ or $\mathrm{W}_{\mathcal{A}_1}^{\mathcal{A}_2}T_{\vartheta}M_{\xi}f(t,u)$)} & $\times\mathrm{e}^{-\mathrm{j}\frac{a_2c_2(a_1-b_1)^2}{2}\vartheta^2}\mathrm{e}^{\mathrm{j}c_2(a_1-b_1)(u-b_2(d_1-c_1)\xi)\vartheta}$\\
 & $\times\mathrm{W}_{\mathcal{A}_1}^{\mathcal{A}_2}f(t-(d_1-c_1)\vartheta,u-a_2(a_1-b_1)\vartheta-b_2(d_1-c_1)\xi)$\\
\specialrule{0.1em}{4pt}{4pt}
\end{tabular}
\begin{tablenotes}
\item[1] The entries in the symplectic matrices should satisfy $a_1=b_1$.
\item[2] The entries in the symplectic matrices should satisfy $a_1=b_1$ and $a_2=d_2=0$.
\item[3] The entries in the symplectic matrices should satisfy $c_1=d_1$ and $a_2=0$.
\end{tablenotes}
\end{threeparttable}
\end{table*}
\subsection{Computational complexity}
\indent The WDL can be regarded as the partial LCT of $\mathfrak{T}_{\begin{pmatrix}1&1\\1/2&-1/2\end{pmatrix}}(f\otimes\overline{f})(t,\varepsilon)$ associated with $\mathcal{A}_2$ with respect to the second variable $\varepsilon$. Since the linear canonical operator $\mathcal{L}^{\mathcal{A}_2}$ maintains the same computational complexity as the Fourier operator $\mathcal{F}$, i.e., $O(M\log M)$ for length $M$ input samples, the computational complexity of the WDL is $O(M^2\log M)$, in accordance with that of the WD.\\
\indent Inspired by the computational complexity analysis of the WD, it is trivial to conclude that the computational complexity of the SWD is identical to that of the WD, i.e., $O(M^2\log M)$ for length $M$ input samples. Therefore, the use of the general symplectic coordinates transformation $\mathfrak{T}_{\mathcal{A}_1\mathbf{J}}$ on the tensor product $f\otimes\overline{f}$ does not increase the computational complexity as compared with the use of the fixed one $\mathfrak{T}_{\begin{pmatrix}1&1\\1/2&-1/2\end{pmatrix}}$. The SWDL can be obtained by generalizing $\mathfrak{T}_{\begin{pmatrix}1&1\\1/2&-1/2\end{pmatrix}}(f\otimes\overline{f})$ found in the WDL to $\mathfrak{T}_{\mathcal{A}_1\mathbf{J}}(f\otimes\overline{f})$. Thus, the SWDL exhibits the same computational complexity as the WDL, i.e., $O(M^2\log M)$.\\
\indent See Table~\ref{tab:3} for a summary of a comparison of computational complexities of the SWDL and others including the WD, WDL and SWD.
\begin{table}[htbp]
\centering
\caption{\label{tab:3}Computational complexities of the WD, WDL, SWD and SWDL}
\footnotesize
\begin{tabular}{ccccc}
\specialrule{0.1em}{4pt}{4pt}
 & WD & WDL & SWD & SWDL\\
\specialrule{0.1em}{4pt}{4pt}
Computational complexity & \multicolumn{4}{c}{$O(M^2\log M)$}\\
\specialrule{0.1em}{4pt}{4pt}
\end{tabular}
\end{table}
\section{Heisenberg's uncertainty principles in the SWDL setting}\label{sec:4}
\indent In this section, we first define the uncertainty product in the SWDL domain and translate it into a summation of two uncertainty products in the LCT domain. We then employ some celebrated Heisenberg's uncertainty principles associated with the uncertainty product in the LCT domain in deducing those in the SWDL setting.
\subsection{Uncertainty product of the SWDL}
\indent\emph{Definition~2:} Let $\mathrm{W}_{\mathcal{A}_1}^{\mathcal{A}_2}f$ be the SWDL of $f\in L^2(\mathbb{R})$ associated with the symplectic matrices $\mathcal{A}_1,\mathcal{A}_2\in Sp(1,\mathbb{R})$. It is then defined as:\\
\indent (i) Spread in the time-SWDL domain:
\begin{equation*}
\vartriangle t_{\mathcal{A}_1,\mathcal{A}_2}^2=\frac{\left\lVert\left(t-t_{\mathcal{A}_1,\mathcal{A}_2}^0\right)\mathrm{W}_{\mathcal{A}_1}^{\mathcal{A}_2}f\right\rVert_2^2}{\left\lVert\mathrm{W}_{\mathcal{A}_1}^{\mathcal{A}_2}f\right\rVert_2^2},
\end{equation*}
where the moment in the time-SWDL domain: $t_{\mathcal{A}_1,\mathcal{A}_2}^0=\left\langle t\mathrm{W}_{\mathcal{A}_1}^{\mathcal{A}_2}f,\mathrm{W}_{\mathcal{A}_1}^{\mathcal{A}_2}f\right\rangle\Big/\left\lVert\mathrm{W}_{\mathcal{A}_1}^{\mathcal{A}_2}f\right\rVert_2^2$.\\
\indent (ii) Spread in the linear canonical frequency-SWDL domain:
\begin{equation*}
\vartriangle u_{\mathcal{A}_1,\mathcal{A}_2}^2=\frac{\left\lVert\left(u-u_{\mathcal{A}_1,\mathcal{A}_2}^0\right)\mathrm{W}_{\mathcal{A}_1}^{\mathcal{A}_2}f\right\rVert_2^2}{\left\lVert\mathrm{W}_{\mathcal{A}_1}^{\mathcal{A}_2}f\right\rVert_2^2},
\end{equation*}
where the moment in the linear canonical frequency-SWDL domain: $u_{\mathcal{A}_1,\mathcal{A}_2}^0=\left\langle u\mathrm{W}_{\mathcal{A}_1}^{\mathcal{A}_2}f,\mathrm{W}_{\mathcal{A}_1}^{\mathcal{A}_2}f\right\rangle\Big/\left\lVert\mathrm{W}_{\mathcal{A}_1}^{\mathcal{A}_2}f\right\rVert_2^2$.\\
\indent\emph{Remark~1:} The spread $\vartriangle t_{\mathcal{A}_1,\mathcal{A}_2}^2$ and the moment $t_{\mathcal{A}_1,\mathcal{A}_2}^0$ in the time-SWDL domain reduce respectively to the spread and the moment in the time-WD domain for $\mathcal{A}_1=\begin{pmatrix}1&1\\-1/2&1/2\end{pmatrix}$ and $\mathcal{A}_2=\mathbf{L}_1$, the spread and the moment in the time-SWD domain for $\mathcal{A}_1$ and $\mathcal{A}_2=\mathbf{L}_1$, and the spread and the moment in the time-WDL domain for $\mathcal{A}_1=\begin{pmatrix}1&1\\-1/2&1/2\end{pmatrix}$ and $\mathcal{A}_2$. The spread $\vartriangle u_{\mathcal{A}_1,\mathcal{A}_2}^2$ and the moment $u_{\mathcal{A}_1,\mathcal{A}_2}^0$ in the linear canonical frequency-SWDL domain reduce respectively to the spread and the moment in the frequency-WD domain for $\mathcal{A}_1=\begin{pmatrix}1&1\\-1/2&1/2\end{pmatrix}$ and $\mathcal{A}_2=\mathbf{L}_1$, the spread and the moment in the frequency-SWD domain for $\mathcal{A}_1$ and $\mathcal{A}_2=\mathbf{L}_1$, and the spread and the moment in the linear canonical frequency-WDL domain for $\mathcal{A}_1=\begin{pmatrix}1&1\\-1/2&1/2\end{pmatrix}$ and $\mathcal{A}_2$.\\
\indent\emph{Remark~2:} Thanks to Parseval's relation of the SWDL, as shown in Eq.~\eqref{eq5}, the spread $\vartriangle t_{\mathcal{A}_1,\mathcal{A}_2}^2$ and the moment $t_{\mathcal{A}_1,\mathcal{A}_2}^0$ in the time-SWDL domain simplify to $\left\lVert\left(t-t_{\mathcal{A}_1,\mathcal{A}_2}^0\right)\mathrm{W}_{\mathcal{A}_1}^{\mathcal{A}_2}f\right\rVert_2^2\Big/\lVert f\rVert_2^4$ and $\left\langle t\mathrm{W}_{\mathcal{A}_1}^{\mathcal{A}_2}f,\mathrm{W}_{\mathcal{A}_1}^{\mathcal{A}_2}f\right\rangle\Big/\lVert f\rVert_2^4$, respectively; the spread $\vartriangle u_{\mathcal{A}_1,\mathcal{A}_2}^2$ and the moment $u_{\mathcal{A}_1,\mathcal{A}_2}^0$ in the linear canonical frequency-SWDL domain simplify to $\left\lVert\left(u-u_{\mathcal{A}_1,\mathcal{A}_2}^0\right)\mathrm{W}_{\mathcal{A}_1}^{\mathcal{A}_2}f\right\rVert_2^2\Big/\lVert f\rVert_2^4$ and $\left\langle u\mathrm{W}_{\mathcal{A}_1}^{\mathcal{A}_2}f,\mathrm{W}_{\mathcal{A}_1}^{\mathcal{A}_2}f\right\rangle\Big/\lVert f\rVert_2^4$, respectively.\\
\indent The uncertainty product in the SWDL domain is defined by the product of the spread $\vartriangle t_{\mathcal{A}_1,\mathcal{A}_2}^2$ in the time-SWDL domain and the spread $\vartriangle u_{\mathcal{A}_1,\mathcal{A}_2}^2$ in the linear canonical frequency-SWDL domain, i.e.,
\begin{equation*}
\vartriangle t_{\mathcal{A}_1,\mathcal{A}_2}^2\vartriangle u_{\mathcal{A}_1,\mathcal{A}_2}^2.
\end{equation*}
In order to simplify the uncertainty product $\vartriangle t_{\mathcal{A}_1,\mathcal{A}_2}^2\vartriangle u_{\mathcal{A}_1,\mathcal{A}_2}^2$ in the SWDL domain, it seems feasible to simplify respectively $\vartriangle t_{\mathcal{A}_1,\mathcal{A}_2}^2$ and $\vartriangle u_{\mathcal{A}_1,\mathcal{A}_2}^2$. Next, we state two preparatory Lemmas which demonstrate an equivalent relationship between $\vartriangle t_{\mathcal{A}_1,\mathcal{A}_2}^2$ and the spread $\vartriangle t^2=\left\lVert\left(t-t^0\right)f\right\rVert_2^2\big/\lVert f\rVert_2^2$ in the time domain and an equivalent relationship between $\vartriangle u_{\mathcal{A}_1,\mathcal{A}_2}^2$ and a summation of two spreads $\vartriangle u_{\mathcal{A}_3}^2=\left\lVert\left(u-u_{\mathcal{A}_3}^0\right)\mathcal{L}^{\mathcal{A}_3}f\right\rVert_2^2\big/\lVert f\rVert_2^2,\vartriangle u_{\mathcal{A}_4}^2=\left\lVert\left(u-u_{\mathcal{A}_4}^0\right)\mathcal{L}^{\mathcal{A}_4}f\right\rVert_2^2\big/\lVert f\rVert_2^2$ in the LCT domain. Here, $t^0=\left\langle tf,f\right\rangle/\lVert f\rVert_2^2$ denotes the moment in the time domain, and $u_{\mathcal{A}_3}^0=\left\langle u\mathcal{L}^{\mathcal{A}_3}f,\mathcal{L}^{\mathcal{A}_3}f\right\rangle\big/\lVert f\rVert_2^2,u_{\mathcal{A}_4}^0=\left\langle u\mathcal{L}^{\mathcal{A}_4}f,\mathcal{L}^{\mathcal{A}_4}f\right\rangle\big/\lVert f\rVert_2^2$ denote moments in the LCT domain.\\
\indent\emph{Lemma~2:} Let $\mathrm{W}_{\mathcal{A}_1}^{\mathcal{A}_2}f$ be the SWDL of $f$ associated with the symplectic matrices $\mathcal{A}_1=\begin{pmatrix}a_1&b_1\\c_1&d_1\end{pmatrix}\in Sp(1,\mathbb{R})$ and $\mathcal{A}_2=\begin{pmatrix}a_2&b_2\\c_2&d_2\end{pmatrix}\in Sp(1,\mathbb{R})$ with $b_2\neq0$, and let $f,tf\in L^2(\mathbb{R})$. There is
\begin{equation}\label{eq6}
\vartriangle t_{\mathcal{A}_1,\mathcal{A}_2}^2=\left(c_1^2+d_1^2\right)\vartriangle t^2.
\end{equation}
\indent\emph{Proof:} From Eq.~\eqref{eq2}, by using the original definition of the SWDL and Parseval's relation of the LCT, the moment $t_{\mathcal{A}_1,\mathcal{A}_2}^0$ in the time-SWDL domain can be expressed as
\begin{align*}
t_{\mathcal{A}_1,\mathcal{A}_2}^0=&\frac{\left\langle t\mathcal{L}^{\mathcal{A}_2}\mathfrak{T}_{\mathcal{A}_1\mathbf{J}}(f\otimes\overline{f}),\mathcal{L}^{\mathcal{A}_2}\mathfrak{T}_{\mathcal{A}_1\mathbf{J}}(f\otimes\overline{f})\right\rangle}{\lVert f\rVert_2^4}\notag\\
\end{align*}
\begin{align*}
=&\frac{\left\langle t\mathfrak{T}_{\mathcal{A}_1\mathbf{J}}(f\otimes\overline{f}),\mathfrak{T}_{\mathcal{A}_1\mathbf{J}}(f\otimes\overline{f})\right\rangle}{\lVert f\rVert_2^4}.
\end{align*}
Taking the change of variables $(t,\varepsilon)\rightarrow(t,\varepsilon)\left(\mathcal{A}_1\mathbf{J}\right)^{-1}=(t,\varepsilon)\begin{pmatrix}-c_1&a_1\\d_1&-b_1\end{pmatrix}$ yields
\begin{align*}
t_{\mathcal{A}_1,\mathcal{A}_2}^0=&\frac{\left\langle(-c_1t+d_1\varepsilon)(f\otimes\overline{f}),f\otimes\overline{f}\right\rangle}{\lVert f\rVert_2^4}\notag\\
=&-c_1\frac{\left\langle tf\otimes\overline{f},f\otimes\overline{f}\right\rangle}{\lVert f\rVert_2^4}+d_1\frac{\left\langle f\otimes\varepsilon\overline{f},f\otimes\overline{f}\right\rangle}{\lVert f\rVert_2^4}\notag\\
=&-c_1\frac{\left\langle tf,f\right\rangle\lVert\overline{f}\rVert_2^2}{\lVert f\rVert_2^4}+d_1\frac{\left\langle\varepsilon\overline{f},\overline{f}\right\rangle\lVert f\rVert_2^2}{\lVert f\rVert_2^4}\notag\\
=&(d_1-c_1)t^0.
\end{align*}
Similarly, the spread $\vartriangle t_{\mathcal{A}_1,\mathcal{A}_2}^2$ in the time-SWDL domain simplifies to
\begin{align*}
\vartriangle t_{\mathcal{A}_1,\mathcal{A}_2}^2=&\frac{\left\lVert\left[t-(d_1-c_1)t^0\right]\mathcal{L}^{\mathcal{A}_2}\mathfrak{T}_{\mathcal{A}_1\mathbf{J}}(f\otimes\overline{f})\right\rVert_2^2}{\lVert f\rVert_2^4}\notag\\
=&\frac{\left\lVert\left[t-(d_1-c_1)t^0\right]\mathfrak{T}_{\mathcal{A}_1\mathbf{J}}(f\otimes\overline{f})\right\rVert_2^2}{\lVert f\rVert_2^4}\notag\\
=&\frac{\left\lVert\left[-c_1\left(t-t^0\right)+d_1\left(\varepsilon-t^0\right)\right](f\otimes\overline{f})\right\rVert_2^2}{\lVert f\rVert_2^4}\notag\\
=&\frac{\left\lVert-c_1\left(t-t^0\right)f\otimes\overline{f}+f\otimes d_1\left(\varepsilon-t^0\right)\overline{f}\right\rVert_2^2}{\lVert f\rVert_2^4}\notag\\
=&c_1^2\frac{\left\lVert\left(t-t^0\right)f\right\rVert_2^2\lVert\overline{f}\rVert_2^2}{\lVert f\rVert_2^4}+d_1^2\frac{\left\lVert\left(\varepsilon-t^0\right)\overline{f}\right\rVert_2^2\lVert f\rVert_2^2}{\lVert f\rVert_2^4}\notag\\
=&\left(c_1^2+d_1^2\right)\vartriangle t^2.
\end{align*}
\qed\\
\indent Lemma~2 indicates that the premultiplication term $\vartriangle t_{\mathcal{A}_1,\mathcal{A}_2}^2$ found in the uncertainty product $\vartriangle t_{\mathcal{A}_1,\mathcal{A}_2}^2\vartriangle u_{\mathcal{A}_1,\mathcal{A}_2}^2$ can be simplified as $\left(c_1^2+d_1^2\right)\vartriangle t^2$.\\
\indent\emph{Lemma~3:} Let $\mathrm{W}_{\mathcal{A}_1}^{\mathcal{A}_2}f$ be the SWDL of $f$ associated with the symplectic matrices $\mathcal{A}_1=\begin{pmatrix}a_1&b_1\\c_1&d_1\end{pmatrix}\in Sp(1,\mathbb{R})$ with $a_1,b_1,c_1,d_1\neq0$ and $\mathcal{A}_2=\begin{pmatrix}a_2&b_2\\c_2&d_2\end{pmatrix}\in Sp(1,\mathbb{R})$ with $b_2\neq0$, let $\mathcal{L}^{\mathcal{A}_3},\mathcal{L}^{\mathcal{A}_4}$ be two linear canonical operators associated with the symplectic matrices $\mathcal{A}_3=\begin{pmatrix}a_2a_1&b_2d_1\\c_2/d_1&d_2/a_1\end{pmatrix},\mathcal{A}_4=\begin{pmatrix}a_2b_1&b_2c_1\\c_2/c_1&d_2/b_1\end{pmatrix}\in Sp(1,\mathbb{R})$, respectively, and let $f,u\mathcal{L}^{\mathcal{A}_3}f,u\mathcal{L}^{\mathcal{A}_4}f\in L^2(\mathbb{R})$. There is
\begin{equation}\label{eq7}
\vartriangle u_{\mathcal{A}_1,\mathcal{A}_2}^2=\vartriangle u_{\mathcal{A}_3}^2+\vartriangle u_{\mathcal{A}_4}^2.
\end{equation}
\indent\emph{Proof:} From Eq.~\eqref{eq3}, by using first the equivalent definition of the SWDL, then Parseval's relation of the FT and finally the change of variable $t\rightarrow b_2c_1d_1t$, the moment $u_{\mathcal{A}_1,\mathcal{A}_2}^0$ in the linear canonical frequency-SWDL domain can be expressed as Eq.~\eqref{eq8}, shown at the top of the next page. By taking the change of variables $(u,v)\rightarrow(u,v)\mathbf{M}^{-1}=(u,v)\begin{pmatrix}1&1\\-1&0\end{pmatrix}$, it follows from Parseval's relation of the LCT that
\begin{figure*}
\begin{align}\label{eq8}
u_{\mathcal{A}_1,\mathcal{A}_2}^0=&|\chi(u,t)|^2\frac{\left\langle u\mathcal{F}\mathfrak{T}_{\mathbf{M}}\left(\widehat{\mathcal{L}^{\mathcal{A}_3}f}\otimes\overline{\widetilde{\mathcal{L}^{\mathcal{A}_4}f}}\right)\left(u,\frac{t}{b_2c_1d_1}\right),\mathcal{F}\mathfrak{T}_{\mathbf{M}}\left(\widehat{\mathcal{L}^{\mathcal{A}_3}f}\otimes\overline{\widetilde{\mathcal{L}^{\mathcal{A}_4}f}}\right)\left(u,\frac{t}{b_2c_1d_1}\right)\right\rangle}{\lVert f\rVert_2^4}\notag\\
=&\frac{1}{|b_2c_1d_1|}\frac{\left\langle u\mathfrak{T}_{\mathbf{M}}\left(\widehat{\mathcal{L}^{\mathcal{A}_3}f}\otimes\overline{\widetilde{\mathcal{L}^{\mathcal{A}_4}f}}\right)\left(u,\frac{t}{b_2c_1d_1}\right),\mathfrak{T}_{\mathbf{M}}\left(\widehat{\mathcal{L}^{\mathcal{A}_3}f}\otimes\overline{\widetilde{\mathcal{L}^{\mathcal{A}_4}f}}\right)\left(u,\frac{t}{b_2c_1d_1}\right)\right\rangle}{\lVert f\rVert_2^4}\notag\\
=&\frac{\left\langle u\mathfrak{T}_{\mathbf{M}}\left(\widehat{\mathcal{L}^{\mathcal{A}_3}f}\otimes\overline{\widetilde{\mathcal{L}^{\mathcal{A}_4}f}}\right),\mathfrak{T}_{\mathbf{M}}\left(\widehat{\mathcal{L}^{\mathcal{A}_3}f}\otimes\overline{\widetilde{\mathcal{L}^{\mathcal{A}_4}f}}\right)\right\rangle}{\lVert f\rVert_2^4}
\end{align}
\hrulefill
\end{figure*}
\begin{align*}
u_{\mathcal{A}_1,\mathcal{A}_2}^0=&\frac{\left\langle(u-v)\left(\widehat{\mathcal{L}^{\mathcal{A}_3}f}\otimes\overline{\widetilde{\mathcal{L}^{\mathcal{A}_4}f}}\right),\widehat{\mathcal{L}^{\mathcal{A}_3}f}\otimes\overline{\widetilde{\mathcal{L}^{\mathcal{A}_4}f}}\right\rangle}{\lVert f\rVert_2^4}\notag\\
=&\frac{\left\langle u\widehat{\mathcal{L}^{\mathcal{A}_3}f}\otimes\overline{\widetilde{\mathcal{L}^{\mathcal{A}_4}f}},\widehat{\mathcal{L}^{\mathcal{A}_3}f}\otimes\overline{\widetilde{\mathcal{L}^{\mathcal{A}_4}f}}\right\rangle}{\lVert f\rVert_2^4}\notag\\
&-\frac{\left\langle\widehat{\mathcal{L}^{\mathcal{A}_3}f}\otimes v\overline{\widetilde{\mathcal{L}^{\mathcal{A}_4}f}},\widehat{\mathcal{L}^{\mathcal{A}_3}f}\otimes\overline{\widetilde{\mathcal{L}^{\mathcal{A}_4}f}}\right\rangle}{\lVert f\rVert_2^4}\notag\\
=&\frac{\left\langle u\widehat{\mathcal{L}^{\mathcal{A}_3}f},\widehat{\mathcal{L}^{\mathcal{A}_3}f}\right\rangle\left\lVert\overline{\widetilde{\mathcal{L}^{\mathcal{A}_4}f}}\right\rVert_2^2}{\lVert f\rVert_2^4}\notag\\
&-\frac{\left\langle v\overline{\widetilde{\mathcal{L}^{\mathcal{A}_4}f}},\overline{\widetilde{\mathcal{L}^{\mathcal{A}_4}f}}\right\rangle\left\lVert\widehat{\mathcal{L}^{\mathcal{A}_3}f}\right\rVert_2^2}{\lVert f\rVert_2^4}\notag\\
=&u_{\mathcal{A}_3}^0-u_{\mathcal{A}_4}^0.
\end{align*}
Similarly, the spread $\vartriangle u_{\mathcal{A}_1,\mathcal{A}_2}^2$ in the linear canonical frequency-SWDL domain simplifies to Eq.~\eqref{eq9}, shown at the top of the next page.\qed\\
\indent Lemma~3 indicates that the postmultiplication term $\vartriangle u_{\mathcal{A}_1,\mathcal{A}_2}^2$ found in the uncertainty product $\vartriangle t_{\mathcal{A}_1,\mathcal{A}_2}^2\vartriangle u_{\mathcal{A}_1,\mathcal{A}_2}^2$ can be simplified as $\vartriangle u_{\mathcal{A}_3}^2+\vartriangle u_{\mathcal{A}_4}^2$.\\
\begin{figure*}
\begin{align}\label{eq9}
\vartriangle u_{\mathcal{A}_1,\mathcal{A}_2}^2=&|\chi(u,t)|^2\frac{\left\lVert\left[u-\left(u_{\mathcal{A}_3}^0-u_{\mathcal{A}_4}^0\right)\right]\mathcal{F}\mathfrak{T}_{\mathbf{M}}\left(\widehat{\mathcal{L}^{\mathcal{A}_3}f}\otimes\overline{\widetilde{\mathcal{L}^{\mathcal{A}_4}f}}\right)\left(u,\frac{t}{b_2c_1d_1}\right)\right\rVert_2^2}{\lVert f\rVert_2^4}\notag\\
=&\frac{1}{|b_2c_1d_1|}\frac{\left\lVert\left[u-\left(u_{\mathcal{A}_3}^0-u_{\mathcal{A}_4}^0\right)\right]\mathfrak{T}_{\mathbf{M}}\left(\widehat{\mathcal{L}^{\mathcal{A}_3}f}\otimes\overline{\widetilde{\mathcal{L}^{\mathcal{A}_4}f}}\right)\left(u,\frac{t}{b_2c_1d_1}\right)\right\rVert_2^2}{\lVert f\rVert_2^4}\notag\\
=&\frac{\left\lVert\left[u-\left(u_{\mathcal{A}_3}^0-u_{\mathcal{A}_4}^0\right)\right]\mathfrak{T}_{\mathbf{M}}\left(\widehat{\mathcal{L}^{\mathcal{A}_3}f}\otimes\overline{\widetilde{\mathcal{L}^{\mathcal{A}_4}f}}\right)\right\rVert_2^2}{\lVert f\rVert_2^4}\notag\\
=&\frac{\left\lVert\left[\left(u-u_{\mathcal{A}_3}^0\right)-\left(v-u_{\mathcal{A}_4}^0\right)\right]\left(\widehat{\mathcal{L}^{\mathcal{A}_3}f}\otimes\overline{\widetilde{\mathcal{L}^{\mathcal{A}_4}f}}\right)\right\rVert_2^2}{\lVert f\rVert_2^4}\notag\\
=&\frac{\left\lVert\left(u-u_{\mathcal{A}_3}^0\right)\widehat{\mathcal{L}^{\mathcal{A}_3}f}\otimes\overline{\widetilde{\mathcal{L}^{\mathcal{A}_4}f}}-\widehat{\mathcal{L}^{\mathcal{A}_3}f}\otimes\left(v-u_{\mathcal{A}_4}^0\right)\overline{\widetilde{\mathcal{L}^{\mathcal{A}_4}f}}\right\rVert_2^2}{\lVert f\rVert_2^4}\notag\\
=&\frac{\left\lVert\left(u-u_{\mathcal{A}_3}^0\right)\widehat{\mathcal{L}^{\mathcal{A}_3}f}\right\rVert_2^2\left\lVert\overline{\widetilde{\mathcal{L}^{\mathcal{A}_4}f}}\right\rVert_2^2}{\lVert f\rVert_2^4}+\frac{\left\lVert\left(v-u_{\mathcal{A}_4}^0\right)\overline{\widetilde{\mathcal{L}^{\mathcal{A}_4}f}}\right\rVert_2^2\left\lVert\widehat{\mathcal{L}^{\mathcal{A}_3}f}\right\rVert_2^2}{\lVert f\rVert_2^4}\notag\\
=&\vartriangle u_{\mathcal{A}_3}^2+\vartriangle u_{\mathcal{A}_4}^2
\end{align}
\hrulefill
\end{figure*}
\indent By combining Lemmas~3 and 4, the uncertainty product $\vartriangle t_{\mathcal{A}_1,\mathcal{A}_2}^2\vartriangle u_{\mathcal{A}_1,\mathcal{A}_2}^2$ can be simplified as $\left(c_1^2+d_1^2\right)\left(\vartriangle t^2\vartriangle u_{\mathcal{A}_3}^2+\vartriangle t^2\vartriangle u_{\mathcal{A}_4}^2\right)$, as given by the following Lemma.\\
\indent\emph{Lemma~4:} Let $\mathrm{W}_{\mathcal{A}_1}^{\mathcal{A}_2}f$ be the SWDL of $f$ associated with the symplectic matrices $\mathcal{A}_1=\begin{pmatrix}a_1&b_1\\c_1&d_1\end{pmatrix}\in Sp(1,\mathbb{R})$ with $a_1,b_1,c_1,d_1\neq0$ and $\mathcal{A}_2=\begin{pmatrix}a_2&b_2\\c_2&d_2\end{pmatrix}\in Sp(1,\mathbb{R})$ with $b_2\neq0$, let $\mathcal{L}^{\mathcal{A}_3},\mathcal{L}^{\mathcal{A}_4}$ be two linear canonical operators associated with the symplectic matrices $\mathcal{A}_3=\begin{pmatrix}a_2a_1&b_2d_1\\c_2/d_1&d_2/a_1\end{pmatrix},\mathcal{A}_4=\begin{pmatrix}a_2b_1&b_2c_1\\c_2/c_1&d_2/b_1\end{pmatrix}\in Sp(1,\mathbb{R})$, respectively, and let $f,tf,u\mathcal{L}^{\mathcal{A}_3}f,u\mathcal{L}^{\mathcal{A}_4}f\in L^2(\mathbb{R})$. There is
\begin{equation}\label{eq10}
\vartriangle t_{\mathcal{A}_1,\mathcal{A}_2}^2\vartriangle u_{\mathcal{A}_1,\mathcal{A}_2}^2=\left(c_1^2+d_1^2\right)\left(\vartriangle t^2\vartriangle u_{\mathcal{A}_3}^2+\vartriangle t^2\vartriangle u_{\mathcal{A}_4}^2\right).
\end{equation}
\indent\emph{Proof:} By multiplying Eqs.~\eqref{eq6} and \eqref{eq7}, we arrive the conclusion.\qed\\
\indent Lemma~4 indicates that the uncertainty product $\vartriangle t_{\mathcal{A}_1,\mathcal{A}_2}^2\vartriangle u_{\mathcal{A}_1,\mathcal{A}_2}^2$ in the SWDL domain equals to a summation of two uncertainty products $\vartriangle t^2\vartriangle u_{\mathcal{A}_3}^2,\vartriangle t^2\vartriangle u_{\mathcal{A}_4}^2$ in the LCT domain, multiplying by $c_1^2+d_1^2$.
\subsection{Lower bounds on the uncertainty product}
\indent Because of Eq.~\eqref{eq10}, the lower bound on the uncertainty product $\vartriangle t_{\mathcal{A}_1,\mathcal{A}_2}^2\vartriangle u_{\mathcal{A}_1,\mathcal{A}_2}^2$ in the SWDL domain can be generated by the lower bound on the uncertainty product $\vartriangle t^2\vartriangle u_{\mathcal{A}}^2$ in the LCT domain. Therefore, it is necessary to recall some well-known Heisenberg's uncertainty principles associated with the uncertainty product $\vartriangle t^2\vartriangle u_{\mathcal{A}}^2$ in the LCT domain, which are summarized in Table~\ref{tab:4}.\\
\indent\emph{Lemma~5 [44], [45]:} Let $\mathcal{F}$ be the Fourier operator, let $\mathcal{L}^{\mathcal{A}}$ be a linear canonical operator associated with the symplectic matrix $\mathcal{A}=\begin{pmatrix}a&b\\c&d\end{pmatrix}\in Sp(1,\mathbb{R})$ with $b\neq0$, and let $f,tf,\omega\mathcal{F}f,u\mathcal{L}^{\mathcal{A}}f\in L^2(\mathbb{R})$. Assume that the classical derivative $f'$ exists at any point $t\in\mathbb{R}$. There is an inequality
\begin{equation}\label{eq11}
\vartriangle t^2\vartriangle u_{\mathcal{A}}^2\geq\frac{b^2}{4}.
\end{equation}
When $f$ is non-zero almost everywhere, the equality holds if and only if $f$ is a Gaussian enveloped complex exponential signal $\mathrm{e}^{-\frac{1}{2\zeta}\left(t-t^0\right)^2+\epsilon}\mathrm{e}^{\mathrm{j}\left(\omega^0t+\varsigma\right)}$, where $\zeta>0$ and $\epsilon,\varsigma\in\mathbb{R}$. Here, $\omega^0=\left\langle\omega\mathcal{F}f,\mathcal{F}f\right\rangle/\lVert f\rVert_2^2$ denotes the moment in the frequency domain.\\
\indent\emph{Lemma~6 [44], [46], [47]:} Let $f$ be a real-valued signal, let $\mathcal{F}$ be the Fourier operator, let $\mathcal{L}^{\mathcal{A}}$ be a linear canonical operator associated with the symplectic matrix $\mathcal{A}=\begin{pmatrix}a&b\\c&d\end{pmatrix}\in Sp(1,\mathbb{R})$ with $b\neq0$, and let $f,tf,\omega\mathcal{F}f,u\mathcal{L}^{\mathcal{A}}f\in L^2(\mathbb{R})$. Assume that the classical derivative $f'$ exists at any point $t\in\mathbb{R}$. There is an inequality
\begin{equation}\label{eq12}
\vartriangle t^2\vartriangle u_{\mathcal{A}}^2\geq\frac{b^2}{4}+\left(a\vartriangle t^2\right)^2.
\end{equation}
When $f$ is non-zero almost everywhere, the equality holds if and only if $f$ is a Gaussian signal $\mathrm{e}^{-\frac{1}{2\zeta}\left(t-t^0\right)^2+\epsilon}$.\\
\indent\emph{Lemma~7 [44], [48]--[50]:} Let $f=\lambda\mathrm{e}^{\mathrm{j}\varphi}$ be a complex-valued signal, let $\mathcal{F}$ be the Fourier operator, let $\mathcal{L}^{\mathcal{A}}$ be a linear canonical operator associated with the symplectic matrix $\mathcal{A}=\begin{pmatrix}a&b\\c&d\end{pmatrix}\in Sp(1,\mathbb{R})$ with $b\neq0$, and let $f,tf,\omega\mathcal{F}f,u\mathcal{L}^{\mathcal{A}}f\in L^2(\mathbb{R})$. Assume that the classical derivative $f'$ exists at any point $t\in\mathbb{R}$. There is an inequality chain
\begin{align}\label{eq13}
\vartriangle t^2\vartriangle u_{\mathcal{A}}^2\geq&\left(\frac{1}{4}+\mathrm{COV}_{t,\omega}^2-\mathrm{Cov}_{t,\omega}^2\right)b^2\notag\\
&+\left(a\vartriangle t^2+b\mathrm{Cov}_{t,\omega}\right)^2\notag\\
\geq&\frac{b^2}{4}+\left(a\vartriangle t^2+b\mathrm{Cov}_{t,\omega}\right)^2,
\end{align}
where $\mathrm{COV}_{t,\omega}=\left\langle\left|t-t^0\right|f,\left|\varphi'-\omega^0\right|f\right\rangle\big/\lVert f\rVert_2^2$ and $\mathrm{Cov}_{t,\omega}=\left\langle\left(t-t^0\right)f,\left(\varphi'-\omega^0\right)f\right\rangle\big/\lVert f\rVert_2^2$ denote the absolute covariance and covariance in the time-frequency domain, respectively. When $f$ is non-zero almost everywhere, the first equality holds if and only if $f$ is a Gaussian enveloped chirp signal with one of the four forms $\mathrm{e}^{-\frac{1}{2\zeta}\left(t-t^0\right)^2+\epsilon}\mathrm{e}^{\mathrm{j}\left[\frac{1}{2\xi}\eta_m(t)\left(t-t^0\right)^2+\omega^0t+\varsigma_m^{\eta_m(t)}\right]}$, $m=1,2,3,4$, where $\xi>0$, $\varsigma_1^1,\varsigma_2^{-1},\varsigma_3^1,\varsigma_3^{-1},\varsigma_4^1,\varsigma_4^{-1}\in\mathbb{R}$, $\eta_1(t)=1$, $\eta_2(t)=-1$, $\eta_3(t)=\mathrm{sgn}\left(t-t^0\right)$ and $\eta_4(t)=-\mathrm{sgn}\left(t-t^0\right)$, (here $\mathrm{sgn}$ denotes the signum operator); the second equality holds if and only if $f$ is a Gaussian enveloped chirp signal with the form of $m=1$ or $m=2$.\\
\indent Combining Lemma 4 with Lemmas~5--7 yields Heisenberg's uncertainty principles associated with the uncertainty product $\vartriangle t_{\mathcal{A}_1,\mathcal{A}_2}^2\vartriangle u_{\mathcal{A}_1,\mathcal{A}_2}^2$ in the SWDL domain, as shown in Table~\ref{tab:5}.\\
\indent\emph{Theorem~1:} Let $\mathrm{W}_{\mathcal{A}_1}^{\mathcal{A}_2}f$ be the SWDL of $f$ associated with the symplectic matrices $\mathcal{A}_1=\begin{pmatrix}a_1&b_1\\c_1&d_1\end{pmatrix}\in Sp(1,\mathbb{R})$ with $a_1,b_1,c_1,d_1\neq0$ and $\mathcal{A}_2=\begin{pmatrix}a_2&b_2\\c_2&d_2\end{pmatrix}\in Sp(1,\mathbb{R})$ with $b_2\neq0$, let $\mathcal{F}$ be the Fourier operator, let $\mathcal{L}^{\mathcal{A}_3},\mathcal{L}^{\mathcal{A}_4}$ be two linear canonical operators associated with the symplectic matrices $\mathcal{A}_3=\begin{pmatrix}a_2a_1&b_2d_1\\c_2/d_1&d_2/a_1\end{pmatrix},\mathcal{A}_4=\begin{pmatrix}a_2b_1&b_2c_1\\c_2/c_1&d_2/b_1\end{pmatrix}\in Sp(1,\mathbb{R})$, respectively, and let $f,tf,\omega\mathcal{F}f,u\mathcal{L}^{\mathcal{A}_3}f,u\mathcal{L}^{\mathcal{A}_4}f\in L^2(\mathbb{R})$. Assume that the classical derivative $f'$ exists at any point $t\in\mathbb{R}$. There are:\\
\indent (i) For arbitrary signals $f$, the inequality
\begin{equation}\label{eq14}
\vartriangle t_{\mathcal{A}_1,\mathcal{A}_2}^2\vartriangle u_{\mathcal{A}_1,\mathcal{A}_2}^2\geq\frac{b_2^2\left(c_1^2+d_1^2\right)^2}{4}
\end{equation}
holds. When $f$ is non-zero almost everywhere, the equality holds if and only if $f$ is a Gaussian enveloped complex exponential signal $\mathrm{e}^{-\frac{1}{2\zeta}\left(t-t^0\right)^2+\epsilon}\mathrm{e}^{\mathrm{j}\left(\omega^0t+\varsigma\right)}$.\\
\indent (ii) For real-valued signals $f$, the inequality
\begin{align}\label{eq15}
\vartriangle t_{\mathcal{A}_1,\mathcal{A}_2}^2\vartriangle u_{\mathcal{A}_1,\mathcal{A}_2}^2\geq&\frac{b_2^2\left(c_1^2+d_1^2\right)^2}{4}\notag\\
&+a_2^2\left(a_1^2+b_1^2\right)\left(c_1^2+d_1^2\right)\left(\vartriangle t^2\right)^2
\end{align}
holds. When $f$ is non-zero almost everywhere, the equality holds if and only if $f$ is a Gaussian signal $\mathrm{e}^{-\frac{1}{2\zeta}\left(t-t^0\right)^2+\epsilon}$.\\
\indent (iii) For complex-valued signals $f$, the inequality chain
\begin{align}\label{eq16}
&\vartriangle t_{\mathcal{A}_1,\mathcal{A}_2}^2\vartriangle u_{\mathcal{A}_1,\mathcal{A}_2}^2\notag\\
\geq&\left(\frac{1}{4}+\mathrm{COV}_{t,\omega}^2-\mathrm{Cov}_{t,\omega}^2\right)b_2^2\left(c_1^2+d_1^2\right)^2\notag\\
&+\Big[\left(a_2a_1\vartriangle t^2+b_2d_1\mathrm{Cov}_{t,\omega}\right)^2\notag\\
&\relphantom{=}+\left(a_2b_1\vartriangle t^2+b_2c_1\mathrm{Cov}_{t,\omega}\right)^2\Big]\left(c_1^2+d_1^2\right)\notag\\
\geq&\frac{b_2^2\left(c_1^2+d_1^2\right)^2}{4}\notag\\
&+\Big[\left(a_2a_1\vartriangle t^2+b_2d_1\mathrm{Cov}_{t,\omega}\right)^2\notag\\
&\relphantom{=}+\left(a_2b_1\vartriangle t^2+b_2c_1\mathrm{Cov}_{t,\omega}\right)^2\Big]\left(c_1^2+d_1^2\right)
\end{align}
holds. When $f$ is non-zero almost everywhere, the first equality holds if and only if $f$ is a Gaussian enveloped chirp signal $\mathrm{e}^{-\frac{1}{2\zeta}\left(t-t^0\right)^2+\epsilon}\mathrm{e}^{\mathrm{j}\left[\frac{1}{2\xi}\eta_m(t)\left(t-t^0\right)^2+\omega^0t+\varsigma_m^{\eta_m(t)}\right]}$, $m\in\{1,2,3,4\}$, and the second equality holds if and only if $f$ is a Gaussian enveloped chirp signal $\mathrm{e}^{-\frac{1}{2\zeta}\left(t-t^0\right)^2+\epsilon}\mathrm{e}^{\mathrm{j}\left[\frac{1}{2\xi}\eta_m(t)\left(t-t^0\right)^2+\omega^0t+\varsigma_m^{\eta_m(t)}\right]}$, $m\in\{1,2\}$.\\
\indent\emph{Proof:} By combining Eq.~\eqref{eq10} with inequalities~\eqref{eq11}--\eqref{eq13}, respectively, we arrive the conclusion.\qed
\begin{table*}[htbp]
\centering
\caption{\label{tab:4}Heisenberg's uncertainty principles associated with the uncertainty product $\vartriangle t^2\vartriangle u_{\mathcal{A}}^2$ in the LCT domain}
\footnotesize
\begin{tabular}{ccc}
\specialrule{0.1em}{4pt}{4pt}
Signal type & Lower bound & Attainable criteria \\
\specialrule{0.1em}{4pt}{4pt}
Arbitrary & $\frac{b^2}{4}$ & Gaussian enveloped complex exponential signal $\mathrm{e}^{-\frac{1}{2\zeta}\left(t-t^0\right)^2+\epsilon}\mathrm{e}^{\mathrm{j}\left(\omega^0t+\varsigma\right)}$ \\
\specialrule{0.1em}{4pt}{4pt}
Real-valued & $\frac{b^2}{4}+\left(a\vartriangle t^2\right)^2$ & Gaussian signal $\mathrm{e}^{-\frac{1}{2\zeta}\left(t-t^0\right)^2+\epsilon}$ \\
\specialrule{0.1em}{4pt}{4pt}
\multirow{2}{*}{Complex-valued} & \multirow{2}{*}{$\frac{b^2}{4}+\left(a\vartriangle t^2+b\mathrm{Cov}_{t,\omega}\right)^2$} & Gaussian enveloped chirp signal \\
& & $\mathrm{e}^{-\frac{1}{2\zeta}\left(t-t^0\right)^2+\epsilon}\mathrm{e}^{\mathrm{j}\left[\frac{1}{2\xi}\eta_m(t)\left(t-t^0\right)^2+\omega^0t+\varsigma_m^{\eta_m(t)}\right]}$, $m\in\{1,2\}$ \\
\specialrule{0.1em}{4pt}{4pt}
\multirow{3}{*}{Complex-valued} & $\left(\frac{1}{4}+\mathrm{COV}_{t,\omega}^2-\mathrm{Cov}_{t,\omega}^2\right)b^2$ & Gaussian enveloped chirp signal \\
& $+\left(a\vartriangle t^2+b\mathrm{Cov}_{t,\omega}\right)^2$ & $\mathrm{e}^{-\frac{1}{2\zeta}\left(t-t^0\right)^2+\epsilon}\mathrm{e}^{\mathrm{j}\left[\frac{1}{2\xi}\eta_m(t)\left(t-t^0\right)^2+\omega^0t+\varsigma_m^{\eta_m(t)}\right]}$, $m\in\{1,2,3,4\}$ \\
\specialrule{0.1em}{4pt}{4pt}
\end{tabular}
\end{table*}
\begin{table*}[htbp]
\centering
\caption{\label{tab:5}Heisenberg's uncertainty principles associated with the uncertainty product $\vartriangle t_{\mathcal{A}_1,\mathcal{A}_2}^2\vartriangle u_{\mathcal{A}_1,\mathcal{A}_2}^2$ in the SWDL domain}
\footnotesize
\begin{tabular}{ccc}
\specialrule{0.1em}{4pt}{4pt}
Signal type & Lower bound & Attainable criteria \\
\specialrule{0.1em}{4pt}{4pt}
Arbitrary & $\frac{b_2^2\left(c_1^2+d_1^2\right)^2}{4}$ & Gaussian enveloped complex exponential signal $\mathrm{e}^{-\frac{1}{2\zeta}\left(t-t^0\right)^2+\epsilon}\mathrm{e}^{\mathrm{j}\left(\omega^0t+\varsigma\right)}$ \\
\specialrule{0.1em}{4pt}{4pt}
Real-valued & $\frac{b_2^2\left(c_1^2+d_1^2\right)^2}{4}+a_2^2\left(a_1^2+b_1^2\right)\left(c_1^2+d_1^2\right)\left(\vartriangle t^2\right)^2$ & Gaussian signal $\mathrm{e}^{-\frac{1}{2\zeta}\left(t-t^0\right)^2+\epsilon}$ \\
\specialrule{0.1em}{4pt}{4pt}
\multirow{3}{*}{Complex-valued} & $\frac{b_2^2\left(c_1^2+d_1^2\right)^2}{4}+\Big[\left(a_2a_1\vartriangle t^2+b_2d_1\mathrm{Cov}_{t,\omega}\right)^2$ & Gaussian enveloped chirp signal \\
& $+\left(a_2b_1\vartriangle t^2+b_2c_1\mathrm{Cov}_{t,\omega}\right)^2\Big]\left(c_1^2+d_1^2\right)$ & $\mathrm{e}^{-\frac{1}{2\zeta}\left(t-t^0\right)^2+\epsilon}\mathrm{e}^{\mathrm{j}\left[\frac{1}{2\xi}\eta_m(t)\left(t-t^0\right)^2+\omega^0t+\varsigma_m^{\eta_m(t)}\right]}$, $m\in\{1,2\}$ \\
\specialrule{0.1em}{4pt}{4pt}
\multirow{4}{*}{Complex-valued} & $\left(\frac{1}{4}+\mathrm{COV}_{t,\omega}^2-\mathrm{Cov}_{t,\omega}^2\right)b_2^2\left(c_1^2+d_1^2\right)^2$ & \multirow{2}{*}{Gaussian enveloped chirp signal} \\
& $+\Big[\left(a_2a_1\vartriangle t^2+b_2d_1\mathrm{Cov}_{t,\omega}\right)^2$ & \multirow{2}{*}{$\mathrm{e}^{-\frac{1}{2\zeta}\left(t-t^0\right)^2+\epsilon}\mathrm{e}^{\mathrm{j}\left[\frac{1}{2\xi}\eta_m(t)\left(t-t^0\right)^2+\omega^0t+\varsigma_m^{\eta_m(t)}\right]}$, $m\in\{1,2,3,4\}$} \\
& $+\left(a_2b_1\vartriangle t^2+b_2c_1\mathrm{Cov}_{t,\omega}\right)^2\Big]\left(c_1^2+d_1^2\right)$ & \\
\specialrule{0.1em}{4pt}{4pt}
\end{tabular}
\end{table*}
\section{Symplectic matrices selection strategy for LFM signal processing}\label{sec:5}
\indent In this section, we investigate the time-frequency resolution of the SWDL, and compare it with time-frequency resolutions of the SWD and WDL. To be specific, we solve the problem of symplectic matrices selection when using the SWDL to deal with LFM signals.
\subsection{The optimal symplectic matrices}
\indent As shown in inequalities~\eqref{eq14}--\eqref{eq16}, only the lower bound $b_2^2\left(c_1^2+d_1^2\right)^2\big/4$ is independent of the signal. The smaller the lower bound, the higher the time-frequency resolution. Thus, the SWDL reaches the highest time-frequency resolution if and only if the lower bound $b_2^2\left(c_1^2+d_1^2\right)^2\big/4$ reaches its minimum value.\\
\indent\emph{Theorem~2:} Given a nonzero constant $b_2$, the optimal symplectic matrices corresponding to the highest time-frequency resolution of the SWDL read $\mathcal{A}_1^{\mathrm{Opt}}=\begin{pmatrix}a_1&b_1\\-b_1/\left(a_1^2+b_1^2\right)&a_1/\left(a_1^2+b_1^2\right)\end{pmatrix}$ and $\mathcal{A}_2^{\mathrm{Opt}}=\mathcal{A}_2=\begin{pmatrix}a_2&b_2\\(a_2d_2-1)/b_2&d_2\end{pmatrix}$.\\
\indent\emph{Proof:} We construct Lagrange function $L(c_1,d_1,\varpi)=c_1^2+d_1^2+\varpi(a_1d_1-b_1c_1-1)$ and set its partial derivatives $\frac{\partial L}{\partial c_1}=2c_1-\varpi b_1$, $\frac{\partial L}{\partial d_1}=2d_1+\varpi a_1$ and $\frac{\partial L}{\partial\varpi}=a_1d_1-b_1c_1-1$ as zero, giving birth to $c_1=-b_1/\left(a_1^2+b_1^2\right)$ and $d_1=a_1/\left(a_1^2+b_1^2\right)$. Thus, the optimal solution of the minimization problem $\min\nolimits_{\mathcal{A}_1,\mathcal{A}_2\in Sp(1,\mathbb{R})}b_2^2\left(c_1^2+d_1^2\right)^2\big/4$ for a given constant $b_2\neq0$ reads $\mathcal{A}_1^{\mathrm{Opt}}=\begin{pmatrix}a_1&b_1\\-b_1/\left(a_1^2+b_1^2\right)&a_1/\left(a_1^2+b_1^2\right)\end{pmatrix}$ and $\mathcal{A}_2^{\mathrm{Opt}}=\mathcal{A}_2=\begin{pmatrix}a_2&b_2\\(a_2d_2-1)/b_2&d_2\end{pmatrix}$.\qed
\subsection{Time-frequency superresolution of the SWDL}
\indent Next, we will demonstrate that the SWDL has a competitive advantage over the SWD and WDL in time-frequency resolution.\\
\indent\emph{Theorem~3:} The time-frequency resolution of the SWDL is higher than that of the SWD if and only if $0<|b_2|<1$, higher than that of the WDL if and only if $0<c_1^2+d_1^2<1/2$, and higher than those of the SWD and WDL if and only if $0<|b_2|<1$ and $0<c_1^2+d_1^2<1/2$.\\
\indent\emph{Proof:} As implied by Table~\ref{tab:1}, the lower bound $b_2^2\left(c_1^2+d_1^2\right)^2\big/4$ reduces to $\left(c_1^2+d_1^2\right)^2\big/4$ for $\mathcal{A}_2=\mathbf{L}_1$, and to $b_2^2/16$ for $\mathcal{A}_1=\begin{pmatrix}1&1\\-1/2&1/2\end{pmatrix}$. Then, the SWDL exhibits higher time-frequency resolution than the SWD if and only if $b_2^2\left(c_1^2+d_1^2\right)^2\big/4<\left(c_1^2+d_1^2\right)^2\big/4$, i.e., $0<|b_2|<1$, and than the WDL if and only if $b_2^2\left(c_1^2+d_1^2\right)^2\big/4<b_2^2/16$, i.e., $0<c_1^2+d_1^2<1/2$. Thus, the SWDL exhibits higher time-frequency resolution than the SWD and WDL if and only if $0<|b_2|<1$ and $0<c_1^2+d_1^2<1/2$.\qed\\
\indent\emph{Theorem~4:} The SWDL associated with the optimal symplectic matrices $\mathcal{A}_1^{\mathrm{Opt}}=\begin{pmatrix}a_1&b_1\\-b_1/\left(a_1^2+b_1^2\right)&a_1/\left(a_1^2+b_1^2\right)\end{pmatrix}$ and $\mathcal{A}_2^{\mathrm{Opt}}=\mathcal{A}_2=\begin{pmatrix}a_2&b_2\\(a_2d_2-1)/b_2&d_2\end{pmatrix}$ satisfying also $a_1^2+b_1^2>2$ and $0<|b_2|<1$ respectively reaches the highest time-frequency resolution which is higher than time-frequency resolutions of the SWD and WDL.\\
\indent\emph{Proof:} Combining Theorems~2 and 3 yields the required result.\qed\\
\indent The above result indicates that the SWDL outperforms the SWD and WDL in energy concentration in time-frequency plane.
\subsection{LFM signal analysis}
\indent The SWDL, inheriting and developing the SWD and WDL, forms its specific superiority in LFM signal processing.\\
\indent\emph{Theorem~5:} The amplitude of the SWDL of a LFM signal $f(t)=\mathrm{e}^{\mathrm{j}\left(\alpha t+\beta t^2\right)}$, $\beta\neq0$ is able to generate an impulse
\begin{equation*}
\sqrt{2\pi|b_2|}\delta\left[u-2\beta(b_1d_1-a_1c_1)b_2t-\alpha(d_1-c_1)b_2\right]
\end{equation*}
if and only if $a_2+2\beta\left(d_1^2-c_1^2\right)b_2=0$.\\
\indent\emph{Proof:} Substituting $f(t)=\mathrm{e}^{\mathrm{j}\left(\alpha t+\beta t^2\right)}$ into Eq.~\eqref{eq2} gives
\begin{align*}
\mathrm{W}_{\mathcal{A}_1}^{\mathcal{A}_2}f(t,u)=&\frac{1}{\sqrt{\mathrm{j}2\pi b_2}}\mathrm{e}^{\mathrm{j}\frac{d_2}{2b_2}u^2}\mathrm{e}^{\mathrm{j}\beta\left(b_1^2-a_1^2\right)t^2}\mathrm{e}^{\mathrm{j}\alpha(b_1-a_1)t}\notag\\
&\times\int_{\mathbb{R}}\mathrm{e}^{\mathrm{j}\left[\frac{a_2}{2b_2}+\beta\left(d_1^2-c_1^2\right)\right]\varepsilon^2}\notag\\
&\relphantom{=}\times\mathrm{e}^{-\mathrm{j}\frac{1}{b_2}\left[u-2\beta(b_1d_1-a_1c_1)b_2t-\alpha(d_1-c_1)b_2\right]}\mathrm{d}\varepsilon,
\end{align*}
and then, it follows that
\begin{align*}
\left|\mathrm{W}_{\mathcal{A}_1}^{\mathcal{A}_2}f(t,u)\right|=&\frac{1}{\sqrt{2\pi|b_2|}}\bigg|\int_{\mathbb{R}}\mathrm{e}^{\mathrm{j}\left[\frac{a_2}{2b_2}+\beta\left(d_1^2-c_1^2\right)\right]\varepsilon^2}\notag\\
&\times\mathrm{e}^{-\mathrm{j}\frac{1}{b_2}\left[u-2\beta(b_1d_1-a_1c_1)b_2t-\alpha(d_1-c_1)b_2\right]}\mathrm{d}\varepsilon\bigg|.
\end{align*}
When $a_2+2\beta\left(d_1^2-c_1^2\right)b_2=0$, we have
\begin{align*}
&\left|\mathrm{W}_{\mathcal{A}_1}^{\mathcal{A}_2}f(t,u)\right|\notag\\
=&\frac{1}{\sqrt{2\pi|b_2|}}\left|\int_{\mathbb{R}}\mathrm{e}^{-\mathrm{j}\frac{1}{b_2}\left[u-2\beta(b_1d_1-a_1c_1)b_2t-\alpha(d_1-c_1)b_2\right]}\mathrm{d}\varepsilon\right|\notag\\
=&\sqrt{2\pi|b_2|}\delta\left[u-2\beta(b_1d_1-a_1c_1)b_2t-\alpha(d_1-c_1)b_2\right].
\end{align*}
\qed\\
\indent By combining Theorems~4 and 5, symplectic matrices of the SWDL for LFM signal processing are selected for $\mathcal{A}_1^{\mathrm{LFM}}=\begin{pmatrix}a_1&b_1\\-b_1/\left(a_1^2+b_1^2\right)&a_1/\left(a_1^2+b_1^2\right)\end{pmatrix}$ and $\mathcal{A}_2^{\mathrm{LFM}}=\begin{pmatrix}2\beta\left(b_1^2-a_1^2\right)b_2\big/\left(a_1^2+b_1^2\right)^2&b_2\\2\beta\left(b_1^2-a_1^2\right)d_2\big/\left(a_1^2+b_1^2\right)^2-1/b_2&d_2\end{pmatrix}$, where $a_1^2+b_1^2>2$ and $0<|b_2|<1$.
\section{Numerical experiments}\label{sec:6}
\indent In this section, we perform a synthesis example to validate the correctness and effectiveness of the derived results.\\
\indent The simulated LFM signal is chosen as $f(t)=\mathrm{e}^{\mathrm{j}\left(t+t^2/2\right)}$, where the initial frequency and frequency rate read $\alpha=1\mathrm{Hz}$ and $\beta=0.5\mathrm{Hz}$, respectively. The observing interval and sampling frequency are set to $[-5\mathrm{s},5\mathrm{s}]$ and $20\mathrm{Hz}$, respectively. Note that in this case the Nyquist-Shannon sampling theorem holds. Fig.~1 conducts a comparison of time-frequency resolutions of the SWDL and some existing methods including the SWD, WDL and WD.\\
\indent Fig.~1(a) plots the contour picture of the SWDL associated with symplectic matrices $\mathcal{A}_1^{\mathrm{LFM}}=\begin{pmatrix}2&2\\-1/4&1/4\end{pmatrix}$ and $\mathcal{A}_2^{\mathrm{LFM}}=\begin{pmatrix}0&1/2\\-2&1\end{pmatrix}$ satisfying $a_1^2+b_1^2=8>2$ and $0<|b_2|=1/2<1$. Fig.~1(c) plots the contour picture of the SWD associated with the symplectic matrix $\mathcal{A}_1^{\mathrm{LFM}}=\begin{pmatrix}2&2\\-1/4&1/4\end{pmatrix}$. Fig.~1(e) plots the contour picture of the WDL associated with the symplectic matrix $\mathcal{A}_2^{\mathrm{LFM}}=\begin{pmatrix}0&1/2\\-2&1\end{pmatrix}$. Fig.~1(g) plots the contour picture of the WD. By employing the Radon transform (RT) in concentrating the pulse energy on the straight line found in the contour picture at the frequency rate $0.5\mathrm{Hz}$ of $f$, it follows the maximum output of the matched filter corresponding to $f$ [23], [24], [35], [38]. Figs.~1(b), (d), (f) and (h) plot frequency rate-amplitude distributions of the RT based SWDL (RT-SWDL), the RT based SWD (RT-SWD), the RT based WDL (RT-WDL) and the RT based WD (RT-WD), respectively. It is clear that the SWDL achieves better time-frequency concentration effect on the LFM signal than the SWD, WDL and WD.
\begin{figure}[htbp]
\centering
\subfigure[\textit{}]
{\label{Fig.sub.1a}\includegraphics[width=0.24\textwidth,height=0.155\textheight]{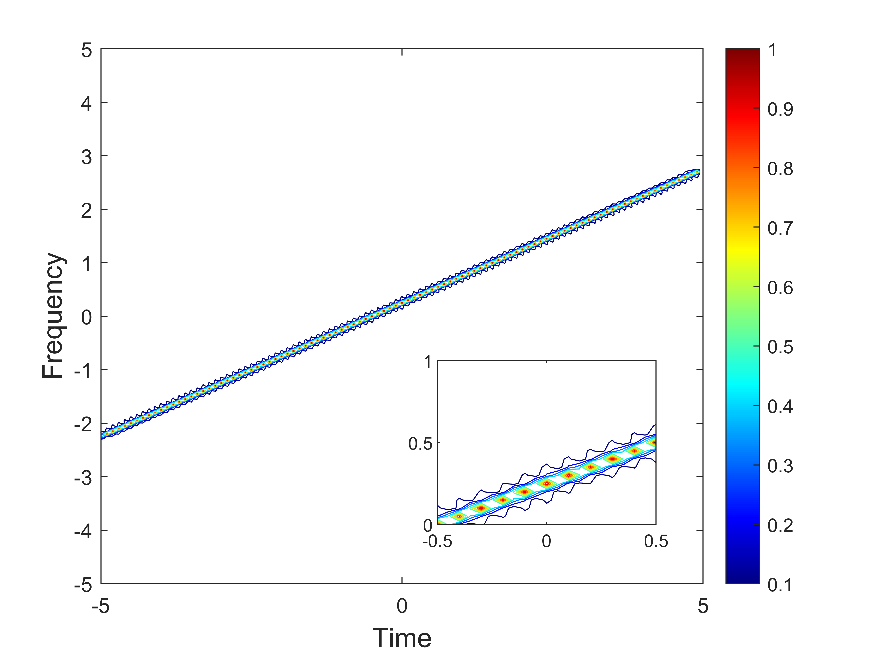}}
\subfigure[\textit{}]
{\label{Fig.sub.1b}\includegraphics[width=0.24\textwidth,height=0.155\textheight]{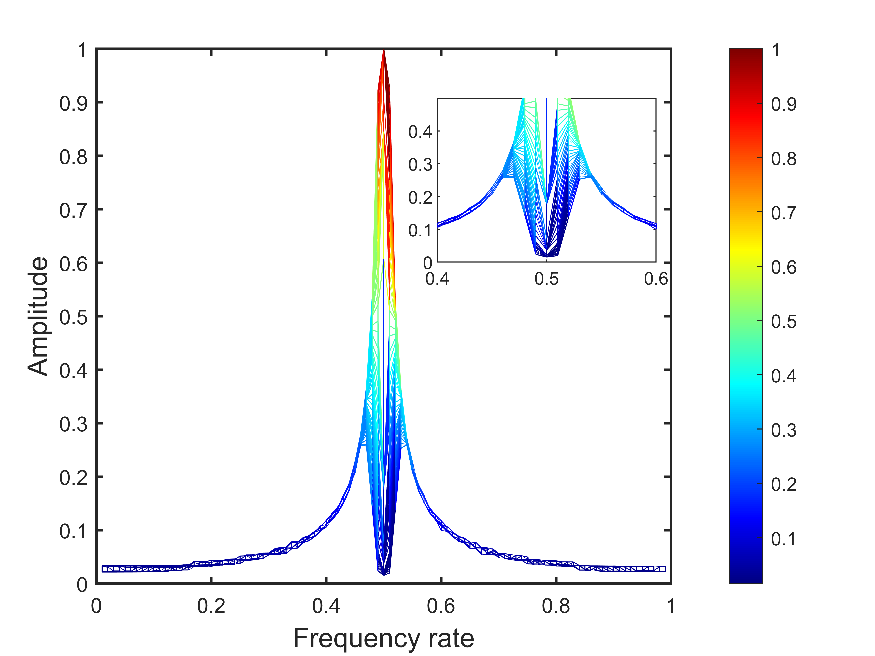}}
\subfigure[\textit{}]
{\label{Fig.sub.1c}\includegraphics[width=0.24\textwidth,height=0.155\textheight]{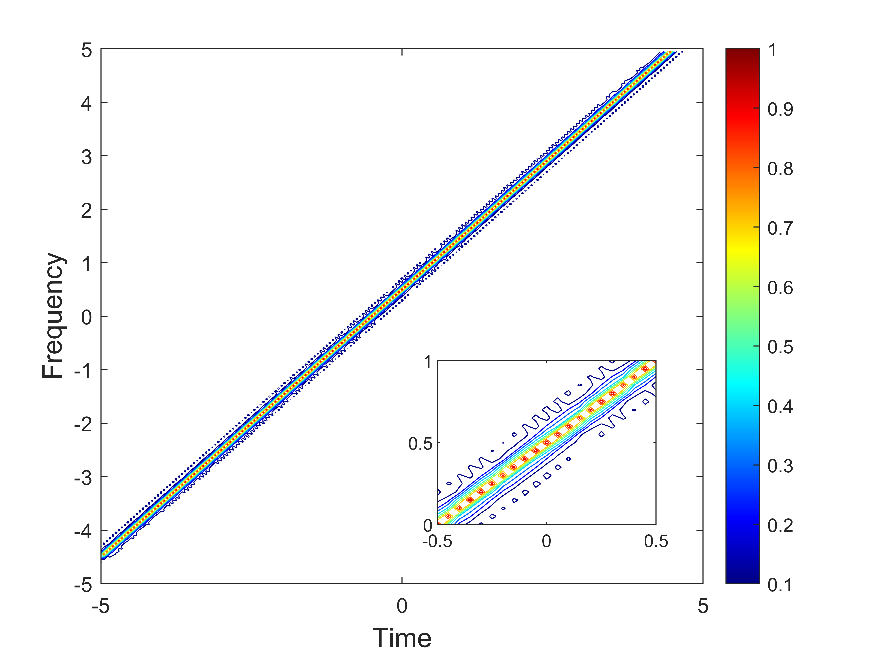}}
\subfigure[\textit{}]
{\label{Fig.sub.1d}\includegraphics[width=0.24\textwidth,height=0.155\textheight]{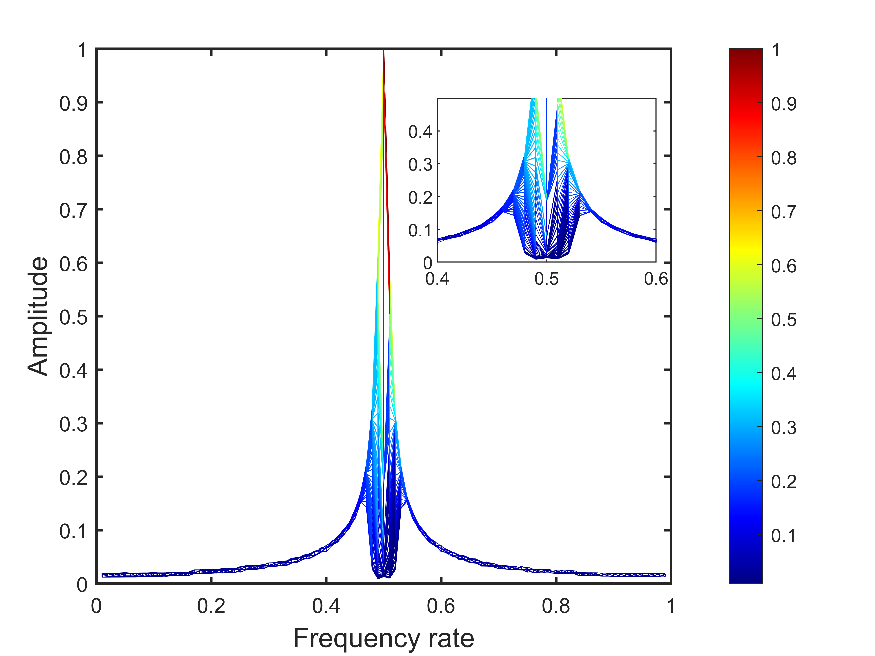}}
\subfigure[\textit{}]
{\label{Fig.sub.1e}\includegraphics[width=0.24\textwidth,height=0.155\textheight]{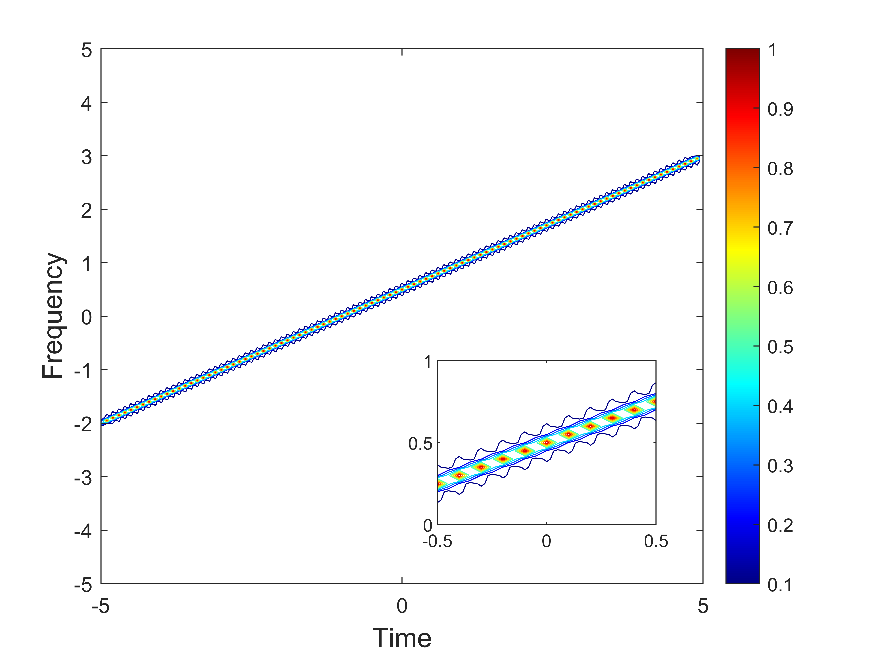}}
\subfigure[\textit{}]
{\label{Fig.sub.1f}\includegraphics[width=0.24\textwidth,height=0.155\textheight]{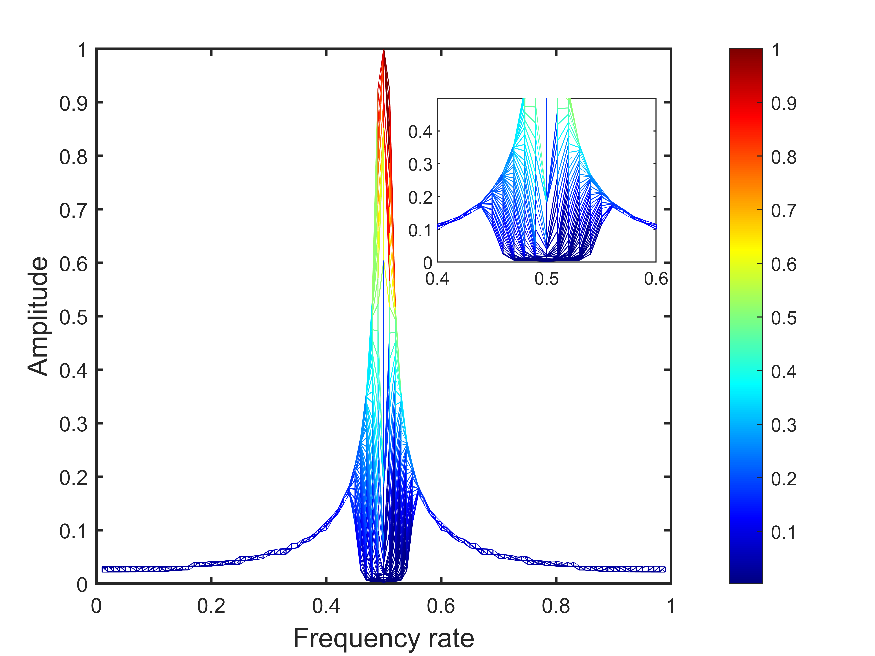}}
\subfigure[\textit{}]
{\label{Fig.sub.1g}\includegraphics[width=0.24\textwidth,height=0.155\textheight]{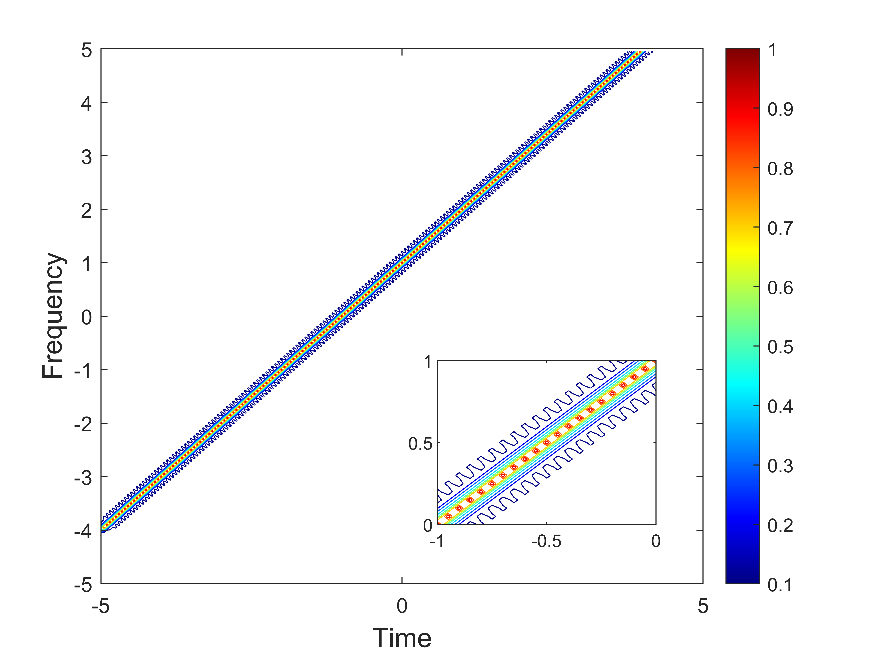}}
\subfigure[\textit{}]
{\label{Fig.sub.1h}\includegraphics[width=0.24\textwidth,height=0.155\textheight]{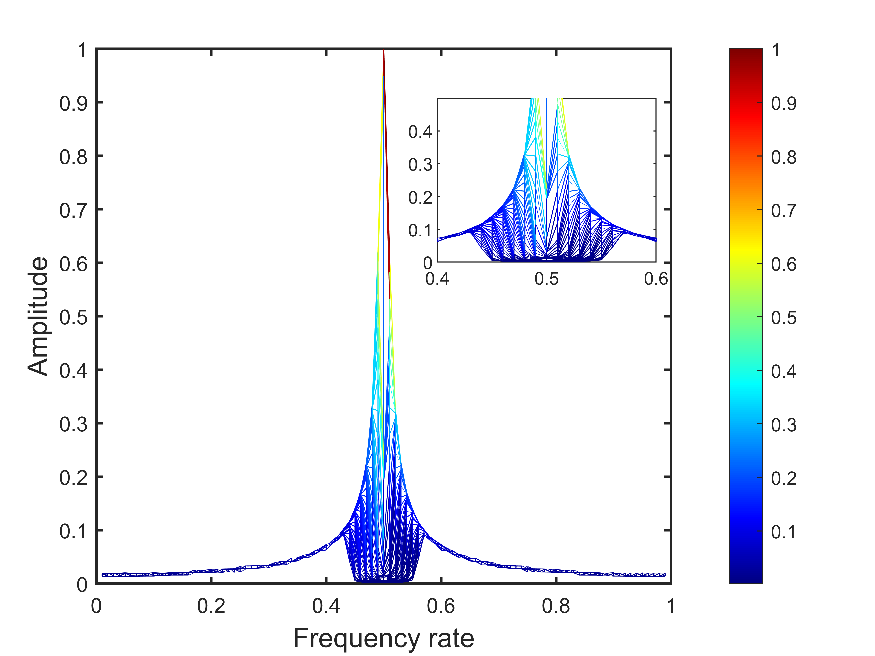}}
\caption{The SWDL, SWD, WDL and WD of the LFM signal. (a) Contour picture of the SWDL; (b) Frequency rate-amplitude distribution of the RT-SWDL; (c) Contour picture of the SWD; (d) Frequency rate-amplitude distribution of the RT-SWD; (e) Contour picture of the WDL; (f) Frequency rate-amplitude distribution of the RT-WDL; (g) Contour picture of the WD; (h) Frequency rate-amplitude distribution of the RT-WD.}
\label{Fig.lable1}
\end{figure}
\section{Conclusion}\label{sec:7}
\indent Time-frequency analysis theory and method of the SWDL that integrates the SWD with WDL have been established. The definition of the SWDL is a direct combination of definitions of the SWD with WDL, providing more flexibility and freedom in non-stationary signal processing without sacrificing the computational complexity. The derived properties of the SWDL, including marginal distributions, energy conservations, unique reconstruction, Moyal formula, complex conjugate symmetry, time reversal symmetry, scaling property, time translation property, frequency modulation property, and time translation and frequency modulation property, are generalizations of those of the SWD, WDL and WD. The obtained Heisenberg's uncertainty principles of the SWDL, accompanied by the lower bound minimization analysis results, demonstrate that the SWDL exhibits higher time-frequency resolution than the SWD with WDL, giving rise to a specific superiority in LFM signals time-frequency energy concentration. These results are not only of theoretical significance, yielding a more useful and effective WD's variant associated with the LCT to extract the frequency rate characteristic of LFM signals, but also of potential application value, especially in radar, communications, sonar, biomedicine and vibration engineering.
\balance

\end{document}